\documentclass[conference]{IEEEtran}
\PassOptionsToPackage{hyphens}{url}\usepackage{hyperref}
\usepackage{tikz}
\usepackage{amsmath,amsfonts,amsthm}
\usepackage{graphicx}
\usepackage{lipsum}
\usepackage{xspace}
\usepackage{algorithm}
\usepackage{listings}
\usepackage{multirow}
\usepackage{marvosym}
\usepackage[english]{babel}
\usepackage[normalem]{ulem}
\usepackage{xurl}
\usepackage{makecell}
\usepackage{circledtext}
\usepackage{pifont}
\usepackage{tabularx}
\usepackage{booktabs}
\usepackage{ifsym}
\newif\ifarxiv
\newcommand{\para}[1]{\noindent{\bf {#1}}}

\newenvironment{packeditemize}{
\begin{list}{$\bullet$}{
\setlength{\itemsep}{1.5pt}
\setlength{\labelwidth}{8pt}
\setlength{\leftmargin}{10pt}
\setlength{\labelsep}{3pt}
\setlength{\listparindent}{\parindent}
\setlength{\parsep}{1.5pt}
\setlength{\parskip}{1.5pt}
\setlength{\topsep}{1.5pt}}}{\end{list}}

\newcommand{\name}{{{DCAT}}\xspace}
\newcommand{\addr}[1]{{\tt #1}}

\begin{document}
\title{Extending Blockchain Untraceability \\with Plausible Deniability}
\author{
\IEEEauthorblockN{Eunchan Park, Kyonghwa Song, Won Hoi Kim, Wonho Song, and Min Suk Kang\textsuperscript{\Letter}}
\IEEEauthorblockA{
KAIST\\
Daejeon, South Korea\\
\{paul.park, song20, wh.kim, swh0329, minsukk\}@kaist.ac.kr
}
}

\IEEEoverridecommandlockouts
\makeatletter\def\@IEEEpubidpullup{6.5\baselineskip}\makeatother

\maketitle
\begin{abstract}
Traditional blockchain untraceability schemes, such as mixers and privacy coins, primarily aim to obscure the sender-receiver relationship of an asset transfer by placing it within an anonymity set.
This paper studies a stronger goal: whether the asset-transfer event itself can be made unobservable by blending into common decentralized-finance (DeFi) activity.
We introduce Deniable Covert Asset Transfer (\name), a class of transfers that deliberately stage common loss-producing events, such as sandwich and arbitrage operations, so that a sender appears to suffer an ordinary economic loss while the receiver appears to earn from that loss. 
We design and validate concrete \name instantiations across two representative blockchain execution environments: a sandwich-based instantiation on Ethereum and an arbitrage-based instantiation on Arbitrum.
Our experiments demonstrate the empirical unobservability of \name under the evaluated settings across both chains.
The evaluated transfers are syntactically identical to corresponding maximal extractable value (MEV) activities, are classified as ordinary extractions by standard MEV detection tools, and leave the sender and receiver unlinked under representative forensic tools.

Since syntactic inspection cannot separate \name from ordinary MEV activity, we examine whether its economic semantics provide useful forensic signals.
Through a large-scale empirical study of MEV losses across Ethereum and Arbitrum, we show that key semantic features exhibit power-law characteristics.
Extreme losses and repeatedly exploited addresses occur in the wild, and thus are not, by themselves, definitive evidence of collusion.
This gives staged transfers plausible deniability and makes fixed-threshold detection prone to false positives.
We therefore develop a multivariate statistical method for forensic triage that ranks incidents by the joint rarity of their economic footprint.
This method narrows a large search space to a manageable set of suspicious cases for manual investigation.
Applying this method to real-world DeFi activity, we showcase three suspicious cases as examples of how such analysis can prioritize incidents for deeper investigation.
\end{abstract}
\section{Introduction}
\label{sec:intro}

Blockchain transactions are inherently transparent.
To maintain a decentralized public ledger, every asset transfer is permanently recorded and globally verifiable, effectively rendering user financial histories fully traceable.
Over the years, numerous works and real-world projects have attempted to overcome this inherent restriction by introducing transaction {\em untraceability} mechanisms to disconnect the link between a sender and a receiver of an asset transfer.
By avoiding sending an asset directly from a sender to a receiver, these mechanisms make a sender and a receiver not directly linkable, thus rendering asset tracing difficult.
Two notable untraceability schemes widely used in the real world are mixers~\cite{deuber2021coinjoin, pertsev2019tornado,bonneau2014mixcoin, valenta2015blindcoin, ruffing2014coinshuffle} and privacy coins~\cite{miers2013zerocoin, sasson2014zerocash,danezis2013pinocchio,garman2014rational, van2013cryptonote}.

The fundamental underpinning of these existing untraceability schemes is to make the true sender-receiver relationship one of many plausible relationships.
Mixers and privacy coins achieve this through different mechanisms, but they share the same forensic goal: constructing a large {\em anonymity set} around an asset transfer.
As a result, forensic analysis must reason over many plausible senders and receivers rather than a single explicit payment path.
One limitation of this anonymity-set approach, however, is that it protects the sender-receiver relationship more directly than the observability of the transfer event itself.
That is, a mixer deposit or a privacy-coin transaction may hide the sender-receiver relationship, but the event remains visible and can still be flagged or prioritized for manual review.
For example, participation in a recognizable mixer (e.g., Tornado Cash~\cite{pertsev2019tornado}) can itself remain visible on chain and serve as a forensic signal.

This paper studies a stronger and qualitatively different goal.
Rather than asking whether the sender and receiver can be linked from visible transfers, we ask whether the asset-transfer event itself can be made indistinguishable from routine decentralized-finance (DeFi) activity.
We call this class of techniques Deniable Covert Asset Transfer, or \name.
In a \name transfer, the sender appears to suffer an ordinary economic loss, while the receiver appears to earn ordinary market profit from that loss.
The economic value moves from the sender to the receiver, but the observable transaction sequence resembles a common market event rather than an explicit asset transfer.

The key insight behind \name is that DeFi systems contain frequent loss-producing events.
Examples include sandwich extractions, arbitrage events, liquidations, and other forms of maximal extractable value (MEV) activity~\cite{qin2022quantifying,daian2020flash,mclaughlin2023large,ferreira2024rolling,luo2026light}.
{For example, a recent work~\cite{luo2026light} reports 1{,}209{,}139 sandwich and 1{,}295{,}829 arbitrage events on Ethereum from January 2023 to December 2024. 
A recent study also reports MEV activity in layer-2 systems such as Arbitrum~\cite{ferreira2024rolling}.}
\name exploits this background activity by deliberately staging such an event between two colluding addresses.
The sender creates a transaction that exposes value to extraction, and the receiver captures the induced loss through a transaction pattern that matches routine market behavior.
As a result, the receiver's gain appears to be MEV profit rather than the endpoint of a covert asset transfer.

Recent public discussions and emerging research suggest that DeFi execution semantics may be abused to move value without leaving a conventional transfer path~\cite{tim2025Money,cointelegraph2025crypto,cryptoalchemy2025defi,cao2026peb}.
These observations raise a broader systems-security question that remains insufficiently understood: when can such value movement become not only hard to {attribute}, but hard to {observe} as an asset transfer at all?
We answer this question by systematizing \name as an unobservability problem in blockchain systems.
Our goal is not to introduce a deployable privacy service, but to expose a forensic blind spot created by the interaction between DeFi market mechanics and {existing graph-based forensic tracing methods}.

We design and demonstrate \name across different economic loss events in layer-1 and layer-2 public blockchain systems.
In particular, we present end-to-end experiments demonstrating sandwich-based \name on Ethereum {Hoodi Testnet} and arbitrage-based \name on Arbitrum {Sepolia testnet}.
Across both chains, the evaluated \name transfers are {\em syntactically identical} to common MEV activities while moving a substantial fraction of the sender's assets to the receiver.
Representative forensic tools fail to reconstruct the sender-receiver relationship because they stop at high-traffic decentralized-exchange contracts and leave the sender and receiver unlinked.
This result demonstrates the {\em empirical unobservability} of \name across two representative blockchain execution environments.

A natural response is then to ask whether \name can be detected {\em semantically} rather than syntactically.
Although a \name event may look like ordinary MEV activity at the transaction-pattern level, its economic intent differs: organic MEV is adversarial and opportunistic, while \name is cooperative and deliberately loss-inducing.
This suggests that \name transfers might be isolated by looking for extreme extraction volumes or repeated losses by the same sender.
However, through a large-scale empirical study of MEV activities across Ethereum and Arbitrum, we show that key semantic features {in our observed dataset} exhibit power-law characteristics across both chains.
Extreme losses and repeatedly exploited addresses occur in the wild, and thus a high-value or severe loss event is not, by itself, definitive evidence of collusion.
A deterministic threshold can either flag many {false positives} or miss carefully tuned \name transfers, giving staged transfers plausible deniability and making fixed-threshold detection brittle.

Nevertheless, this limitation does not preclude forensic triage.
For investigations where the stakes justify manual review, such as large-scale illicit fund tracing, the practical goal is often not fully automated attribution but effective prioritization.
To this end, we develop a multivariate statistical method for ranking suspicious incidents.
The method models the joint distribution of multiple semantic features and ranks incidents by how statistically rare their combined economic footprint is under the baseline of MEV activity in the wild.
Investigators can use the resulting ranking to narrow a large search space to a manageable set of suspicious cases for manual review.
We apply this method to real-world DeFi activity and showcase three suspicious cases that illustrate how our multivariate analysis can surface incidents that merit deeper investigation.

This paper makes the following contributions.
\begin{packeditemize}
    \item We systematize \name as an unobservability problem in blockchain systems (\S\ref{sec:dcat}).
    \name hides economic value movement inside routine loss-producing DeFi activity rather than inside an explicit anonymity set of transfers.

    \item We design and validate concrete \name instantiations across two execution environments (\S\ref{sec:empirical-validation}): a sandwich-based \name and an arbitrage-based \name on Ethereum and Arbitrum testnets, respectively.
    We show empirical unobservability under the evaluated settings across both chains.

    \item We conduct a large-scale empirical study of MEV losses (\S\ref{sec:semantic-inspection}).
    We show that key semantic features exhibit power-law characteristics, which gives extreme staged losses plausible deniability and limits simple threshold-based detection.

    \item We develop a multivariate statistical method for forensic triage that prioritizes suspicious incidents for manual review by modeling the joint rarity of their economic footprint (\S\ref{sec:multivariate-detection}).
    We showcase three real-world cases surfaced by this analysis.
\end{packeditemize}
\section{Background and Related Work}
\label{sec:background}

This section provides the technical background and related work needed to position \name.
We first review existing blockchain untraceability schemes, focusing on how mixers and privacy coins rely on anonymity sets to obscure sender-receiver relationships~(\S\ref{sec:existing-untraceability}).
We then summarize blockchain forensic techniques and their reliance on explicit transaction-flow reconstruction~(\S\ref{sec:existing-forensic}).
Finally, we discuss prior public speculation and emerging research on MEV-assisted value movement, clarifying how our work differs by systematizing \name as an unobservability problem and empirically studying its feasibility, forensic limitations, and detection prospects~(\S\ref{sec:mev-assisted-laundery}).

\subsection{Untraceability via Anonymity Sets}
\label{sec:existing-untraceability}

Given the {inherent transparency property} of blockchains, many untraceability schemes~\cite{deuber2021coinjoin,pertsev2019tornado,bonneau2014mixcoin, valenta2015blindcoin,ruffing2014coinshuffle,miers2013zerocoin, sasson2014zerocash,moser2013inquiry,alonso2020zero} have been {proposed} to {obscure relationships among protocol participants}.
A core goal of these schemes is to prevent an observer from confidently linking a sender to the corresponding receiver after inspecting on-chain data~\cite{pfitzmann2001anonymity}.
Most {existing} schemes pursue this goal by constructing \emph{anonymity sets} over candidate senders and receivers.
That is, even when candidate participants are known, forensic analysis should not be able to identify which sender corresponds to which receiver with high confidence. % given a particular sender or vice versa.
We next review two prevalent classes of such schemes and how they construct anonymity sets.

Blockchain mixers, such as CoinJoin~\cite{deuber2021coinjoin} and Tornado Cash~\cite{pertsev2019tornado},
pool funds from multiple users and later distribute them to designated recipient addresses, creating an anonymity set of participants.
This pooled redistribution obscures the direct relationship between each sender and its intended receiver.
{Depending on the design, mixers may operate through a centralized service~\cite{bonneau2014mixcoin, valenta2015blindcoin} or a decentralized protocol~\cite{pertsev2019tornado, ruffing2014coinshuffle}, and may support fixed denominations~\cite{deuber2021coinjoin, wang2023zero} or variable deposit amounts~\cite{wu2021towards}.}
Regardless of these design choices, their common objective is to make multiple sender-receiver pairings plausible within the pool, thereby weakening direct transaction-flow tracing.

{Privacy coins~\cite{miers2013zerocoin, sasson2014zerocash,danezis2013pinocchio,garman2014rational, van2013cryptonote} embed anonymity mechanisms directly into the blockchain protocol, providing anonymity sets at the protocol level.}
Zerocoin~\cite{miers2013zerocoin} and its successors~\cite{sasson2014zerocash,danezis2013pinocchio,garman2014rational} use zero-knowledge proofs to decouple coin deposit from later spending, ensuring that no observer can link the two events.
Monero~\cite{van2013cryptonote} uses ring signatures to conceal the true sender-receiver relationship.
These mechanisms construct anonymity sets by design, making participants indistinguishable within a set of plausible senders or receivers.

\subsection{Transaction-Graph Forensics}
\label{sec:existing-forensic}
In response to the growing use of blockchain for illicit activity~\cite{turner2018bitcoin, farrugia2020detection}, forensic techniques have evolved {along two directions, each tailored to the UTXO-based and account-based blockchains.}

First, UTXO-based blockchain forensics dates back to 2013, when several influential works~\cite{meiklejohn2013fistful, androulaki2013evaluating, reid2012analysis, ron2013quantitative} introduced two foundational heuristics for tracing Bitcoin transactions: (1) the co-spending heuristic posits that all input addresses of a transaction are controlled by the same entity, and (2) the shadow address heuristic assumes that the change output of a UTXO transaction is returned to an address owned by the same entity as the inputs.
{By applying these, prior works constructed address relationship graphs that cluster addresses controlled by the same entity.
Subsequent research extended these foundations by refining clustering heuristics, enriching transaction graphs with contextual role annotations~\cite{gomez2022watch, spagnuolo2014bitiodine}, validating forensic methodologies through law-enforcement practice~\cite{lubbertsen2025ghost}, and developing specialized frameworks for tracing illicit fund flows such as ransomware payments~\cite{huang2018tracking,liao2016behind}. 
More recently, Kappos et al.~\cite{kappos2022peel} further improved clustering by combining machine learning with structural transaction patterns such as peel chains.}

Second, the account-based model {employed by} blockchains such as Ethereum {introduces structural differences} from UTXO-based designs, {leading to corresponding adaptations in forensic methodology.}
{Unlike Bitcoin forensics, Ethereum tracing lacks a de facto standard heuristic; existing approaches instead follow explicit transaction flows and apply targeted heuristics to reduce false positives. For instance, prior work has proposed heuristics for high-confidence address clustering~\cite{victor2020address}, score-based relationship graphs for source attribution~\cite{lin2024denseflow}, taint-analysis frameworks for tracing illicit flows in real-world heists~\cite{wu2023toward}, and biased graph-search techniques that incorporate transaction-level details into tracing~\cite{wu2023tracer}.
}

\subsection{Prior Discussions on MEV-Assisted Value Movement}
\label{sec:mev-assisted-laundery}
{
Prior to this work, there existed public speculation regarding the use of MEV extraction as a vehicle for covert asset transfer. Following a massive MEV extraction event in March 2025, several articles~\cite{tim2025Money,cointelegraph2025crypto,cryptoalchemy2025defi} raised suspicions that the extraction may have constituted money laundering disguised as MEV activity. The incident prompted debate among industry experts, including personnel from Cyvers, Hacken, DefiLlama, MetaMask, Flashbots, and Uniswap. Nevertheless, no formal follow-up investigation was conducted to determine whether such large-scale extractions indeed represented laundering behavior, leaving these claims unsubstantiated.

More recently, Cao et al.~\cite{cao2026peb} published an arXiv preprint that independently discusses a related form of covert value movement through DeFi execution semantics.
The preprint articulates a mechanism closely related to the transfer technique studied in this paper.
At the same time, however, it addresses a narrower scope than the systems-security questions we study here.
In particular, the preprint focuses primarily on modeling, leaving open broader security questions about the underlying unobservability properties, practical feasibility, and forensic detectability of such transfers.

Our work builds on this emerging line of discussion and provides a systems-security study of MEV-assisted covert asset transfer.
We systematize \name as an unobservability problem, validate concrete instantiations across Ethereum and Arbitrum, analyze why simple semantic detection is limited, and develop an advanced forensic-triage method for prioritizing suspicious real-world cases.
}
\section{Deniable Covert Asset Transfer}
\label{sec:dcat}
Unlike conventional untraceability schemes surveyed in the previous section, which rely on anonymity sets, \name takes a different approach: it hides asset movement inside a staged loss event that is common in the DeFi ecosystem.
In this section, we first formalize the adversary model in which \name operates~(\S\ref{sec:threat-model}), develop the core intuition through a concrete toy example~(\S\ref{sec:toy-example}), and discuss practical factors that affect real-world feasibility~(\S\ref{sec:details-sandwich-based}).
\subsection{Threat Model}
\label{sec:threat-model}
For the purpose of this paper, we focus on the adversarial setting of \name, where a malicious actor seeks to transfer assets in an untraceable manner by exploiting on-chain economic loss events. 
Note, however, that \name can also be utilized in a non-adversarial context, such as by users seeking to enhance their privacy without malicious intent.
Under this adversary model, we define the adversary's capabilities and objectives as follows.

The \name adversary seeks to transfer a target digital asset from a sender address to a receiver address in an {\em untraceable manner}, without relying on well-known mechanisms such as mixers and privacy coins.
This motivation is practical: existing privacy mechanisms are often recognizable onchain and have received substantial regulatory and forensic scrutiny~\cite{ustreasury22}. 
An adversary may therefore seek a transfer mechanism that does not appear as participation in a known untraceability service. 

We consider the adversary to have no capabilities beyond those available to any benign blockchain participant: they may submit transactions to the network, interact with arbitrary smart contracts, and optionally participate in the consensus mechanism of the underlying chain.
The adversary does not possess privileged access to any protocol component, network infrastructure, or any vulnerability in the system that is unavailable to ordinary users.

In addition to the target asset held by the sender, the adversary may control separate working capital at the receiver address.
This working capital is not the value being covertly transferred, but a reserve used to execute the staged market action that captures the sender's induced loss.
We assume that this reserve is operationally separate from the target asset, so the observable outcome remains a sender-side loss and a receiver-side market profit rather than a direct transfer.

\subsection{High-level Intuition with a Toy Example}
\label{sec:toy-example}

\begin{figure}[t]
    \centering
    \includegraphics[width=0.5\textwidth]{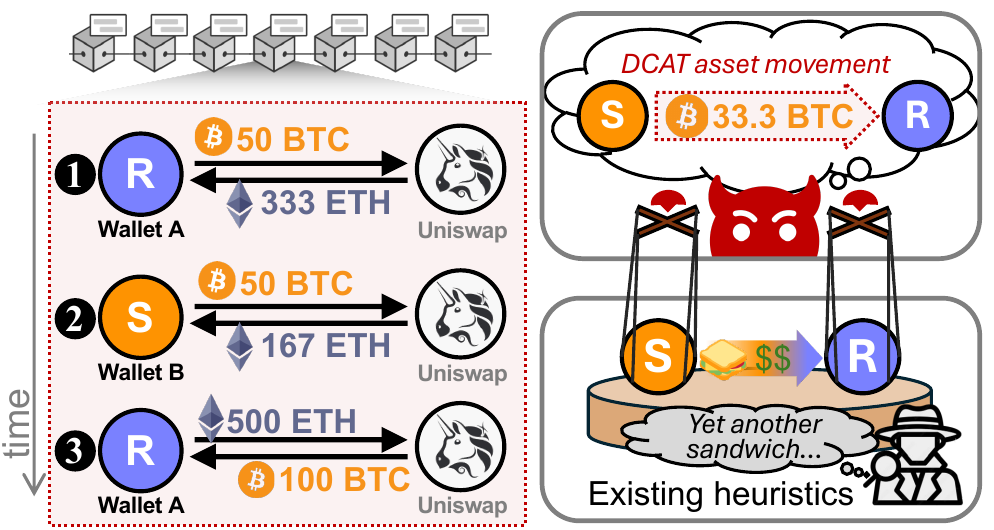}
    \caption{A toy example of the sandwich-based \name attack. The \name sender (S) and receiver (R) {\em stage} a sandwich extraction on the Uniswap DEX (\ding{182}--\ding{184}). Existing heuristics are limited to interpreting it as yet another sandwich operation (thus failing to recognize it as a deliberate transfer) while a significant amount of assets (33.3~BTC) is transferred from \addr{S} to \addr{R} in an {\em unobservable} manner.}
    \label{fig:toy-example}
\end{figure}

\para{A toy example.}
We illustrate the core intuition of \name using the toy example in Figure~\ref{fig:toy-example}.
Consider an adversary who controls two addresses, \addr{S} and \addr{R}, and wishes to move digital assets from \addr{S} to \addr{R} without creating an observable transfer event.
In this toy example (and throughout the remainder of the paper), we use a {\em sandwich operation} as the running example of an economic loss event because it is a common MEV activity within the current DeFi ecosystem~\cite{qin2022quantifying}.

\circledtextset{resize=real}

Let us walk through the sequence of events illustrated in Figure~\ref{fig:toy-example}, where \addr{S} operates as the {\em \name sender} and \addr{R} serves as the {\em \name receiver}. 
The core of the \name transfer is the highlighted three-transaction sandwich sequence. 
Prior to this sandwich operation, address \addr{S} holds 50~BTC, and address \addr{R} holds 50~BTC and 167~ETH. 
The adversary controls both addresses and intends to move part of the value held by \addr{S} to \addr{R}.
The adversary is aware that \addr{S}'s transaction history is publicly recorded and may later be use for forensic tracing. 
The goal is therefore to move value without leaving a traceable link from \addr{S} to \addr{R}.

To initiate the process, the \name adversary uses \addr{S} to execute a standard token swap (\ding{183}) on a decentralized exchange (DEX), such as Uniswap, attempting to exchange BTC for ETH. 
Such a swap is fully transparent to the public and thus does not offer any inherent untraceability.

The obfuscation occurs when the adversary uses the designated address \addr{R} as a {\em sandwicher} to execute a staged sandwich operation targeting the {\em sandwichee} address \addr{S}'s Uniswap transaction.
To achieve this, \addr{R} submits a {\em frontrunning} transaction (\ding{182}) immediately before \addr{S}'s swap (\ding{183}) and a {\em backrunning} transaction (\ding{184}) immediately after.
The frontrunning transaction purchases ETH ahead of \addr{S}, inflating its exchange rate; \addr{S}'s swap then executes at this unfavorable rate, resulting in a significant loss; finally, the backrunning transaction sells the ETH acquired in the frontrun at the now-elevated price, capturing the profit.
This standard sandwich operation deliberately manipulates the local exchange rate, allowing \addr{R} to extract a profit directly from the financial loss incurred by \addr{S}.
After the staged sandwich concludes, \addr{R} holds the extracted profit as ordinary MEV revenue.

\para{How much is transferred?}
The {\em transferred amount} corresponds to the profit retrieved by the receiver from the staged loss. 
In our toy example, the adversary aims to maximize the proportion of \addr{S}'s loss that is captured by \addr{R}. 
By deliberately configuring transaction parameters, most notably by setting a high slippage tolerance on \addr{S}'s swap, the adversary ensures that the induced loss is maximized.
As a result of this orchestrated sequence, \addr{S} incurs a loss of 33.3~BTC, while \addr{R}'s balance increases by 33.3~BTC, assuming a stable market rate of 1~BTC = 10~ETH.
That is, this \name operation successfully transfers 33.3~BTC from \addr{S} to \addr{R}, which is 66.7\% of \addr{S}'s original 50~BTC.
This toy example reports the gross transferred amount, abstracting away operational costs such as transaction fees for simplicity.
In practice, these costs slightly reduce the receiver's net gain and therefore the transfer effectiveness, but the reduction is expected to be marginal when the transferred value is sufficiently large.

\para{Unobservability intuition.}
We now explain why the staged sequence is difficult to recognize as an asset transfer under existing tracing heuristics.
When a forensic tool traces from \addr{S}, it fails to cluster \addr{S} with \addr{R} because existing tools typically treat high-traffic smart contracts such as DEXs as terminating points of each exchange transaction, rather than as intermediaries that facilitate asset transfer.
At the same time, the transaction sequence matches the structure of a standard sandwich operation.
As a result, existing tools are likely to interpret the sequence as ordinary MEV activity rather than as a deliberate transfer from \addr{S} to \addr{R}.

\subsection{From the Toy Example to Real-World Feasibility}
\label{sec:details-sandwich-based}
Although the toy example captures the core idea, realizing it in practice requires careful operational planning.
We discuss three factors that an adversary should consider for effective execution of a sandwich-based \name transfer and briefly survey the landscape of exploitable DEX swap pools in the wild.
\begin{packeditemize}
\item \emph{DEX types.}
To achieve a \name transfer with a significant amount, the adversary would need to choose a DEX that is particularly susceptible to exchange rate manipulation, as the profit margin of a sandwich operation is directly tied to the degree of price impact it can induce.
Uniswap is one of the most widely used DEXs today~\cite{uniswap-share}, with an average of 14.1 million monthly active users as of 2025~\cite{uniswap-stats}. 
Uniswap V2 is particularly well-suited as its constant-product formula ($x \cdot y = k$) distributes liquidity uniformly across the entire price curve, producing a nonlinear impact on the exchange rate. 
For example, a swap that induces a \$10 loss at \$100 input can exceed \$100 in loss when scaled to \$1,000.
The toy example~(\S\ref{sec:toy-example}) therefore uses Uniswap V2-style constant-product DEX.
\item \emph{Capital and pool liquidity.}
The ratio of swap input to pool liquidity governs the degree of exchange rate manipulation, so the adversary must estimate the expected loss for a given capital deployment prior to execution.
A critical constraint is that exchange rate inflation must not exceed the point at which the backrunning transaction can no longer recover most of the induced loss. 
From our toy example, if the adversary cannot fund the additional 167~ETH for \addr{R} and can only backrun with the swapped 333~ETH, it can only receive 80~BTC instead of 100~BTC, leaving 20~BTC in the pool.
Within this constraint, effectiveness scales monotonically with capital deployment.
\item \emph{Single vs. multiple transfers.}
An adversary could transfer a large lump sum in a single \name transfer, as shown in the toy example, but this may not always be feasible due to capital constraints. 
In such cases, the adversary can use multiple \name transfers, each transferring a portion of the total amount, to achieve the overall transfer goal.
\end{packeditemize}

\para{Exploitable DEX swap pools in the wild.}
To assess real-world feasibility, we surveyed available Uniswap V2 pools via Dune Analytics~\cite{dune-analytics} {and identified approximately 4{,}800 pools that could be exploited for the sandwich-based \name attack in the toy example}.
{
The number of exploitable pools depends on the adversary's available capital.
For example, an adversary deploying approximately \$2{,}000 (\$900 for the sender and \$1{,}100 for the receiver) to transfer roughly \$600 in a single \name transfer can exploit approximately 2{,}000 of these pools.
As the adversary's capital increases, the number of exploitable pools grows accordingly.
}
\section{Empirical Validation of Unobservability}
\label{sec:empirical-validation}

We now empirically validate that \name can be executed end-to-end and can reproduce the observable structure of ordinary MEV activity. 
Our goal is to test the two properties that underlie \name's {\em unobservability}: syntactic identity with organic MEV incidents and evasion of existing graph-based forensic tracing.
We first define these mechanics~(\S\ref{sec:unobservability}), and then validate them across two execution environments:  
a sandwich-based transfer on Ethereum Hoodi Testnet~(\S\ref{sec:demo-ethereum}) and an arbitrage-based transfer on Arbitrum Sepolia testnet~(\S\ref{sec:demo-arbitrum}).

\subsection{Mechanics of Unobservability}
\label{sec:unobservability}
To define the privacy guarantees of \name, we ground our evaluation in the established taxonomy introduced by Pfitzmann and Köhntopp~\cite{pfitzmann2001anonymity}, who define unobservability as ``the state of items of interest (IOIs) being indistinguishable from any IOI at all.'' 
In our threat model, the primary IOI is the deliberate, covert asset transfer between a sender and a receiver. 
\name does not merely obscure the identities behind a known transfer; instead, it aims to make the occurrence of the transfer itself indistinguishable from routine, non-transfer ecosystem activity. 
This state of unobservability is achieved through the intersection of two distinct characteristics: the \emph{syntactic identity} of the staged transactions and the \emph{structural blind spots} inherent in modern forensic methodologies.

\para{Indistinguishability via \emph{syntactic identity}.}
Syntactic identity serves as the fundamental mechanism that enables this indistinguishability {from the routine organic extractions in the ecosystem}. 
\name orchestrates economic loss events that carefully mirror the transaction types, ordering, and state changes of organic MEV extractions, such as sandwich attacks or arbitrages. 
By strictly adhering to the expected structural footprint of common market loss activities, \name avoids triggering mechanized anomaly detections. 
Consequently, when automated detection systems or analysts observe the event, they classify \name transfers as routine economic events, establishing the \emph{plausible deniability} necessary to mask the underlying IOI.

\para{Structural \emph{blind spots} in forensics.}
{\name's camouflage} may still fail if forensic entities can exhaustively trace all interconnected asset flows used in the \name transfer. 
\name secures its unobservability by deliberately operating within the structural blind spots of existing forensic clustering algorithms. 
Existing forensic tools categorically exclude high-traffic smart contracts, such as DEXs, from their address clusters to prevent massive false-positive explosions.
{For this practical reason, they choose} to treat these contracts as terminating endpoints rather than shared intermediaries. 
\name exploits this limitation by operating the covert transfer through these untraceable boundaries.

\para{Achieving unobservability.}
{Consequently}, it is the combination of these two characteristics that renders the \name transfer unobservable. 
Syntactic identity ensures that the transaction is immediately classified as background MEV noise, preventing initial suspicion. 
Concurrently, the structural blind spot {ensures} that forensic tools {would} halt their clustering at the DEX contract, ensuring that the hidden multi-hop transfer is not reconstructed. 
Because the event appears entirely {ordinary} on the surface and occurs exactly where forensics overlook, the deliberate asset transfer remains indistinguishable from the absence of a transfer, matching our operational criterion for empirical unobservability.

\subsection{Validation on Ethereum: Sandwich-based \name}
\label{sec:demo-ethereum}

We now empirically validate the syntactic identity and forensic blind spots by executing an end-to-end \name transfer in a live environment, beginning with a sandwich-based instantiation on Ethereum.
Our validation proceeds in three steps: we first construct the transfer, then confirm that a widely used detection tool classifies it as an organic sandwich, and finally show that existing forensic tools fail to link the sender and receiver.

To create \name transfers syntactically identical to organic sandwich incidents, we demonstrate it with the \emph{same pipeline} through which organic sandwiches are produced: the Proposer-Builder Separation (PBS) architecture~\cite{PBS-ethereum}. 
PBS distributes block construction across four roles.
Searchers identify profitable sandwich opportunities and submit them as transaction bundles to builders; builders assemble these bundles into complete blocks, attach a bid, and forward the block to a relay; the relay forwards the highest-bid block to the assigned validator, who proposes it and receives the corresponding bid; refer to~\cite{PBS-ethereum} for a detailed description of the PBS ecosystem.

To stage a sandwich operation that flows through this pipeline, a \name adversary {should} operate its own searcher and builder.
Both roles are inexpensive to operate, as open-source implementations are publicly available, and neither requires specialized hardware or staked capital. 

The adversary may additionally operate its own validator; while not strictly required, doing so improves reliability and eliminates the dominant operational cost of the transfer.
{
This is because, without controlling the block-proposing validator, the builders in PBS have to compete in a bid auction and pay the bid to the proposing validator to propose its block; this bid constitutes an operational cost for the adversary.  
When the validator is under adversarial control, however, the bid simply returns to the adversary, allowing its builder to submit an arbitrarily high bid and guarantee inclusion.
}

The only remaining task is to wait for the adversarial validator to be assigned a proposal slot and to ensure the builder submits its bid within the 12-second block time. 
This opportunistic aspect of the attack is acceptable when the adversary can wait for an assigned proposal slot, given the unobservability benefit and high transfer effectiveness of the \name transfer.

\para{Testbed environment and execution.}
We conducted our experiments on the Hoodi testnet, which provides a live but isolated Ethereum environment {for experiments}.
For the block builder, we used Flashbots' rbuilder~1.3.7~\cite{rbuilder-github}, configured to connect to a public Flashbots relay~\cite{flashbot-relay-hoodi}.
For the validator, we used Lighthouse~8.1.0~\cite{lighthouse-github} as the consensus client, Reth~1.11.0~\cite{reth-github} as the execution client, and MEV-Boost~\cite{mev-boost-github} configured to connect to the same relay.
{For an identical setting with the mainnet}, we deployed a liquidity pool for a custom ERC-20 token pair using a fork of the Uniswap~V2 codebase~\cite{uniswap-github}.

{We then ran a custom Python-based searcher that monitors} our validator's scheduled proposal slot and orchestrates the transfer as follows.
When the adversary's validator is scheduled to propose the next block, the adversary submits the sandwichee's (i.e., the \name sender's) transaction to the public mempool.
The searcher simultaneously constructs a \name sandwich bundle, syntactically identical to a standard sandwich, and submits it to the block builder.
Because the colluding validator recaptures the bid, the adversary's builder submits an arbitrarily high bid to guarantee inclusion, completing the covert transfer while preserving the outward appearance of an ordinary sandwich incident.

\para{Validating syntactic identity.}
We have successfully executed a \name transfer in the Hoodi testnet, and {\em our \name transfer} is visible in the explorer{\footnote{\url{https://hoodi.etherscan.io/tx/0xb56889b4cefdfd32c30e57050ec6f492bc7b65fe5ef40c540542b11f6489a2cc}}\footnote{\url{https://hoodi.etherscan.io/tx/0x0da364ef963b2820c38fe20799a3d9b785d1a66e12f13e8dce3277babe7125cb}}\footnote{\url{https://hoodi.etherscan.io/tx/0x6d59cdb4f251447f47c1414159d35716ac9e0e565306f3afc180dc3a961b8e93}}}.
To confirm that our \name transfer is syntactically identical to an organic sandwich at the transaction level, we analyzed the crafted incident using a widely used open-source sandwich detection tool~\cite{mev-inspect-github}. 
This tool flags {a well-known} three-transaction pattern in which the first and third transactions originate from the same address, all three interact with the same DEX, and the first and second swap the same token pair while the third swaps in the reverse direction.
In our executed incident, our sandwicher \addr{0x11} and the sandwichee \addr{0xEA} interacted with our custom Uniswap pair \addr{0x3E}, exactly reproducing this structural pattern.
The detection tool {successfully} classified our \name incident as an ordinary sandwich, empirically confirming that the required syntactic structure was replicated.

\para{Validating evasion of forensics.}
{
Confirming the syntactic identity, we now empirically validate the unobservability~(\S\ref{sec:unobservability}) of the orchestrated \name transfer demonstrated in the network. 
For this purpose, we applied two {representative} open-source Ethereum forensic tools~\cite{victor2020address,wu2023tracer} to the \name transfer.
}
Both tools either failed to link the \name sender to the receiver because they excluded the shared DEX contract, or produced a massive false-positive explosion when attempting to cluster through the shared DEX.
{
These outcomes highlight a fundamental limitation inherent to existing forensic tools.
To the best of our knowledge, existing graph-based forensic methods do not provide a reliable way to trace through high-traffic DEX contracts without incurring a false-positive explosion.
}
This suggests that \name resides within a systematic blind spot of the evaluated forensic tools and remains unlinked under the evaluated tracing methodologies.

\subsection{Validation on Arbitrum: Arbitrage-based \name}
\label{sec:demo-arbitrum}

{ 
We next validate the same unobservability mechanism on the Arbitrum layer-2 execution model.
Unlike Ethereum, Arbitrum uses a centralized sequencer that orders transactions on a first-come-first-served basis, which renders standard frontrunning challenging, making arbitrage the dominant extraction strategy instead, as reported earlier by Ferreira et al.~\cite{ferreira2024rolling}.
Consequently, the \name adversary adapts its staged loss to an arbitrage, mimicking an organic arbitrageur who exploits a sudden price discrepancy. 
}

{
An arbitrage-based \name transfer is produced by two sequential transactions.
First, the \name sender submits a rate-inflating swap that intentionally skews the exchange rate of a target pool.
The \name receiver then submits a backrunning arbitrage transaction that captures the staged loss before another non-colluding arbitrageur can intervene.
A critical constraint is that these two transactions must land in \emph{different} blocks.
This is because, as Arbitrum has no public mempool, arbitrageurs can observe the inflated exchange rate only after the rate-inflating transaction is published on-chain; consequently, a rate-inflating transaction and an arbitrage transaction appearing in the same block would itself constitute observable evidence of collusion. 
The adversary must therefore time its submissions precisely: submitting the arbitrage transaction too early risks co-inclusion with the rate-inflating swap, while submitting too late risks preemption by a competing arbitrageur.
}

\para{Testbed environment and execution.}
{
To demonstrate that this timing constraint can be satisfied in practice, we deployed our experimental setup on the Arbitrum Sepolia testnet.
We used two AWS EC2 instances to submit transactions: one in Ohio to submit \name transactions with minimal latency to the Arbitrum sequencer in Ohio, and another in North Virginia to send {probing transactions} that help estimate block boundaries with high precision.
Similar to the Ethereum setup, we deployed a liquidity pool for a custom ERC-20 token pair using a fork of Uniswap~V2~\cite{uniswap-github}. 
}

{
To reliably estimate the block boundary, we consistently sent probing transactions disguised as normal transfers, initiating the \name transfer only upon confirming the exact boundary. 
This strict timing ensures that the rate-inflating transaction lands at the very end of one block, allowing the arbitrage transaction to execute in subsequent blocks. 
As a result, this probing strategy achieves a 98.3\% success rate in a practical setup; we defer the fine-grained timing analysis to Appendix~\ref{sec:arbitrum-probing}. 
}

\para{Validating syntactic identity.}
{
We have conducted the arbitrage-based \name transfer in the Arbitrum Sepolia testnet, and both of {\em our rate-inflating and arbitrage transactions} are visible in the explorer\footnote{\url{https://sepolia.arbiscan.io/tx/0x82811aa91ab862cc7afec0ebf52825de05a01df7923caa6b5040e0022f5bb3b7}}\footnote{\url{https://sepolia.arbiscan.io/tx/0x737f4641e4522b5327a8b22c03703bc9ba77593b0b787adde1b6c90edb7c94df}}.
Using the arbitrage-based \name conducted in the network, we now verify the \name transfer's syntactic identity by analyzing our crafted incident using a standard open-source arbitrage detection tool~\cite{mev-inspect-github}.
This tool identifies arbitrage events by tracing swap histories for an arbitrage transaction that forms a closed cycle of asset movements\footnote{A closed-cycle design allows the arbitrageur to hold assets in a single asset type, simplifying arbitrage across multiple asset types. This is a common practice since the contract can be reused without changing any arguments.} following a rate-inflating transaction.
In our execution, the receiver \addr{0xCD} executed an arbitrage swap against the inflated pool \addr{0x25}, captured the staged loss incurred by the sender \addr{0x07} two blocks earlier, and swapped back to the original asset type.
The detection tool accordingly classified our incident as a standard arbitrage event, confirming that the expected syntactic structure was replicated.
}

\para{Validating evasion of forensics.}
{
We also test our own \name transfers in the Arbitrum Sepolia testnet against available forensic tools. 
Similar to our sandwich-based \name transfer, the forensic tracing tools~\cite{victor2020address,wu2023tracer} again either halted their cluster expansion at the shared DEX contract to avoid a false-positive explosion or resulted in a false-positive explosion by clustering the shared DEX, leaving the sender and receiver unlinked. 

By reproducing this outcome across distinct layer-1 and layer-2 execution models, we show that \name's unobservability mechanism is not tied to a single chain architecture.
}
\section{Semantic Divergence and Forensic Limitations}
\label{sec:semantic-inspection}
{In the previous section}, we showed that \name transfers can reproduce the syntactic structure of organic market activity. 
This syntactic identity, however, does not imply {\em semantic equivalence}.
{Organic MEV extractions} are adversarial and {opportunistic} while \name transfers are {cooperative and deliberately loss-inducing.}
This section {examines whether this semantic difference} can support forensic inspection.
We first introduce the idea of isolating \name transfers by analyzing the economic behavior of transacting entities~(\S\ref{sec:intuition-for-semantic-inspection}).
{We then build empirical baselines from large-scale MEV activity on Ethereum and Arbitrum~(\S\ref{sec:empirical-baselines}), showing that the relevant extraction features exhibit power-law characteristics.
Finally, we explain why these heavy-tailed baselines make fixed-threshold inspection brittle and provide plausible deniability for staged losses~(\S\ref{sec:limitations}).

\subsection{Intuition for Threshold-based Semantic Inspection}
\label{sec:intuition-for-semantic-inspection}

The high-level intuition behind {semantic inspection} is that \name and organic MEV activity, while syntactically similar, differ in their economic intent.
In organic MEV, extraction is adversarial and opportunistic. 
An extractee inadvertently exposes value to extraction, which an independent extractor captures. 
Thus, the organic extractee may have incentives to reduce such losses, for example, by setting tight slippage tolerance~\cite{zhou2021high} or by avoiding repeated exposure from the same address.

In contrast, \name relies on cooperation between the sender and receiver. 
The \name sender deliberately weakens these guardrails to induce a large loss that the receiver can capture as apparent MEV profit.
This suggests a simple semantic inspection strategy: identify unusually large losses, unusually high extraction ratios, or repeated extraction patterns that deviate from ordinary MEV behavior.
To test whether such threshold-based inspection is viable, we conduct a large-scale empirical study of MEV extractions in the wild.

\subsection{Semantic Features in Underlying MEV Activities}
\label{sec:empirical-baselines}
{
To assess the practical feasibility of {semantic inspection}, we conducted a large-scale measurement study of MEV activities occurring on layer-1 Ethereum and layer-2 Arbitrum.
Our analysis focuses on four features that capture the economic magnitude and long-term operational patterns of each MEV activity:
\begin{packeditemize}
\item {\em [F1] Extraction volume}: 
The absolute profit captured by an extractor. This corresponds to the transferred amount in a \name transfer.
\item {\em [F2] Capital-extraction ratio}: 
An extractee's loss expressed as a fraction of its swapped capital.
This reflects the severity of the extractee's guardrail bypass.
\item {\em [F3] Bilateral extraction frequency}: 
The number of incidents involving the same extractor-extractee pair.
A high frequency might indicate repeated interaction between the same two addresses.
\item {\em [F4] Extractee exploitation frequency}: 
The total number of times a specific extractee address is targeted across all extractors.
A high frequency may indicate an address that repeatedly exposes value to extraction.
\end{packeditemize}
We selected these four features because they explicitly quantify the economic guardrails discussed previously.
To transfer a substantial amount of value, an adversary may need to drive some of these metrics to large values, thereby possibly diverging from the baseline behavior of ordinary users.
}

\para{Data collection.}
We establish {an empirical statistical baseline} of extraction activities from two representative MEV ecosystems: {sandwich and arbitrage exploitations on Ethereum and Arbitrum, respectively}.
We gathered sandwich incidents on Ethereum using Mev-inspect-py~\cite{mev-inspect-github} and arbitrage incidents on Arbitrum using the open-source code published by Ferreira et al.~\cite{ferreira2024rolling,rolling-github}.
For each incident, we reconstruct the profit and loss in USD at the moment of extraction, using prices obtained from CoinGecko~\cite{coingecko}.
Our data collection spans March 8, 2025 to December 13, 2025 (block numbers 22{,}000{,}000 to 23{,}999{,}999) for Ethereum and March 1, 2025 to April 20, 2025 {(block numbers 310{,}947{,}140 to 327{,}902{,}139)} for Arbitrum.
{From the collected dataset, we} exclude incidents involving cryptocurrencies for which CoinGecko does not provide reliable price data {(e.g., meme coins with extreme price volatility or assets lacking USD price feeds)}, accounting for approximately 43.1\% of incidents on Ethereum and 63.3\% on Arbitrum.
Our data covers a total 313{,}219 sandwich incidents on Ethereum and 55{,}482 arbitrage incidents on Arbitrum.

\para{Highlights: {\em power-law} characteristics.}
Our primary question is whether large losses and repeated extraction patterns are rare enough to serve as reliable forensic separators.
If ordinary MEV activity already contains heavy-tailed extraction behavior, then a staged \name transfer can remain plausible even when it appears economically severe.

Our results show that the relevant semantic features exhibit power-law characteristics.
For unbounded features such as [F1] extraction volume, [F3] bilateral extraction frequency, and [F4] extractee exploitation frequency, a power-law tail means that increasingly large observations remain part of the same scale-free empirical pattern rather than forming a clearly separated regime.
For the bounded [F2] capital-extraction ratio, the analogous behavior appears near the upper boundary: values can approach full extraction closely, and the observed tail shows that near-boundary losses also occur in the baseline.

We present these characteristics for all four features on Ethereum and Arbitrum.
Although the fitted tail behavior varies across features and chains, the common pattern is that simple magnitude-based thresholds do not provide a clean separation between ordinary MEV activity and staged losses.

A canonical visual indicator of power-law characteristics is an approximately straight line on a log-log Complementary Cumulative Distribution Function (CCDF).
This visual signature is shown across all Figures~\ref{fig:extraction-volume}--\ref{fig:extractee-exploitation-frequency}.
{To better quantify the power-law characteristics in the tail, we evaluate the \emph{tail exponent ($\alpha$)}, which characterizes the rate of tail decay (smaller is heavier), with empirical power laws typically lying in $(2, 3)$~\cite{clauset2009power,newman2005power}.}

\begin{figure}[t]
    \centering
    \includegraphics[width=0.45\textwidth]{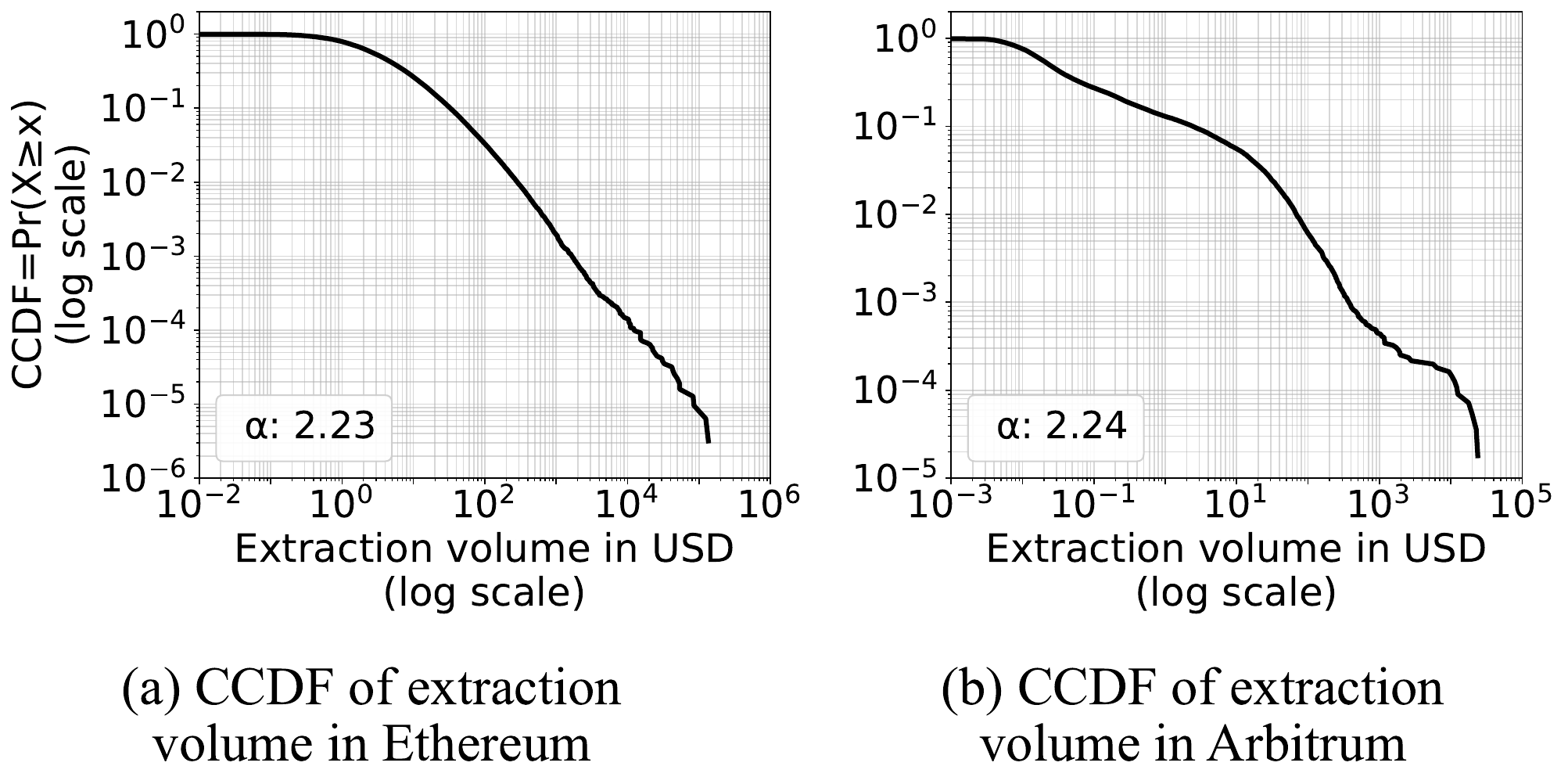}
    \caption{CCDF of the [F1] extraction volume on Ethereum (a) and Arbitrum (b).}
    \label{fig:extraction-volume}
\end{figure}

\para{[F1] Extraction volume.}
As shown in Figure~\ref{fig:extraction-volume}, the CCDF of extraction volume is approximately linear on a log-log scale on both chains, exhibiting the canonical visual signature of a power law.
The tail exponents ($\alpha = 2.23$ on Ethereum, $\alpha = 2.24$ on Arbitrum) sit well within the empirical $(2, 3)$ range, indicating a tail heavy enough that the theoretical variance diverges.
A heavy-tailed extraction-volume baseline weakens magnitude-only separation of high-value \name transfers; that is, because {observed} incidents produce five- or {even} six-figure extractions, a staged \name transfer of comparable size is not automatically separable by magnitude alone.

\begin{figure}[t]
    \centering
    \includegraphics[width=0.45\textwidth]{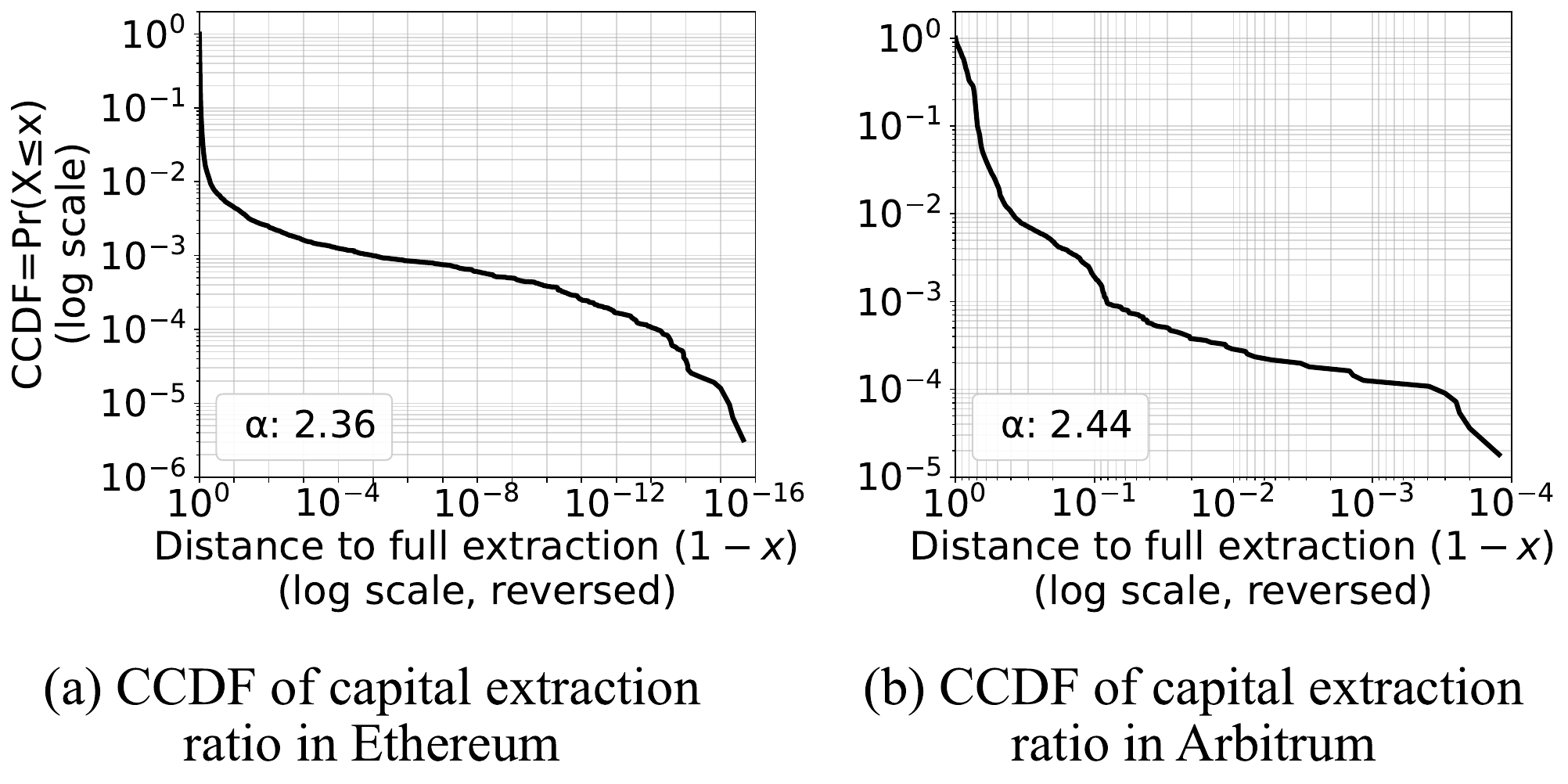}
    \caption{CCDF of the [F2] capital-extraction ratio on Ethereum (a) and Arbitrum (b).}
    \label{fig:capital-extraction-ratio}
\end{figure}

\para{[F2] Capital-extraction ratio.}
\label{sec:capital-extraction-ratio}
{
The capital-extraction ratio also exhibits power-law characteristics.
Although this ratio is inherently bounded within the range of $[0, 1)$ (since an extractee cannot lose more than the initial capital and some non-zero fees are always incurred), the ratio can be infinitely close to one. 
Figure~\ref{fig:capital-extraction-ratio} measures how close the ratio gets to one in terms of its distance to one (i.e., $1 - \text{capital-extraction ratio}$). 
We observe the tail trend in both chains in the CCDF plot and calculate scaling parameters ($\alpha = 2.36$ and $\alpha = 2.44$ on Ethereum and Arbitrum, respectively).
This suggests that near-full extraction ratios are part of the observed tail rather than isolated artifacts.
}

\begin{figure}[t]
    \centering
    \includegraphics[width=0.45\textwidth]{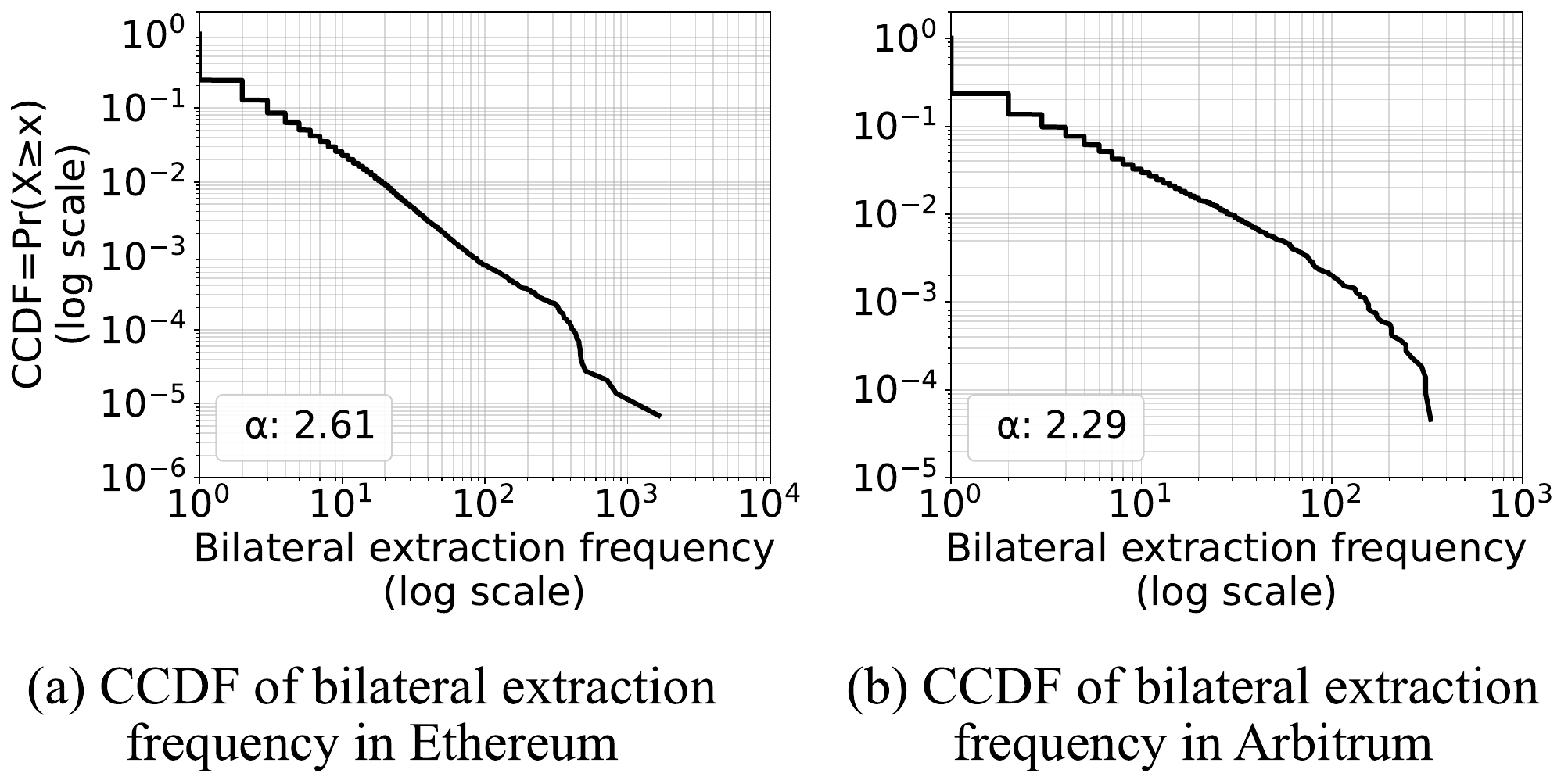}
    \caption{CCDF of [F3] bilateral extraction frequency on Ethereum (a) and Arbitrum (b).}
    \label{fig:bilateral-extraction-frequency}
\end{figure}

\para{[F3] Bilateral extraction frequency.}
\label{sec:bilateral-extraction-frequency}
Figure~\ref{fig:bilateral-extraction-frequency} again exhibits the linear log-log CCDF characteristic of a power law, with fitted exponents of $\alpha = 2.61$ on Ethereum and $\alpha = 2.29$ on Arbitrum, both within the canonical heavy-tail range.
For \name, this heavy tail means that recurring transfers between a fixed sender--receiver pair are not anomalous by frequency alone: extractors in the wild routinely revisit the same victim dozens or even hundreds of times, providing cover for repeated \name campaigns between a fixed pair of addresses.

\begin{figure}[t]
    \centering
    \includegraphics[width=0.45\textwidth]{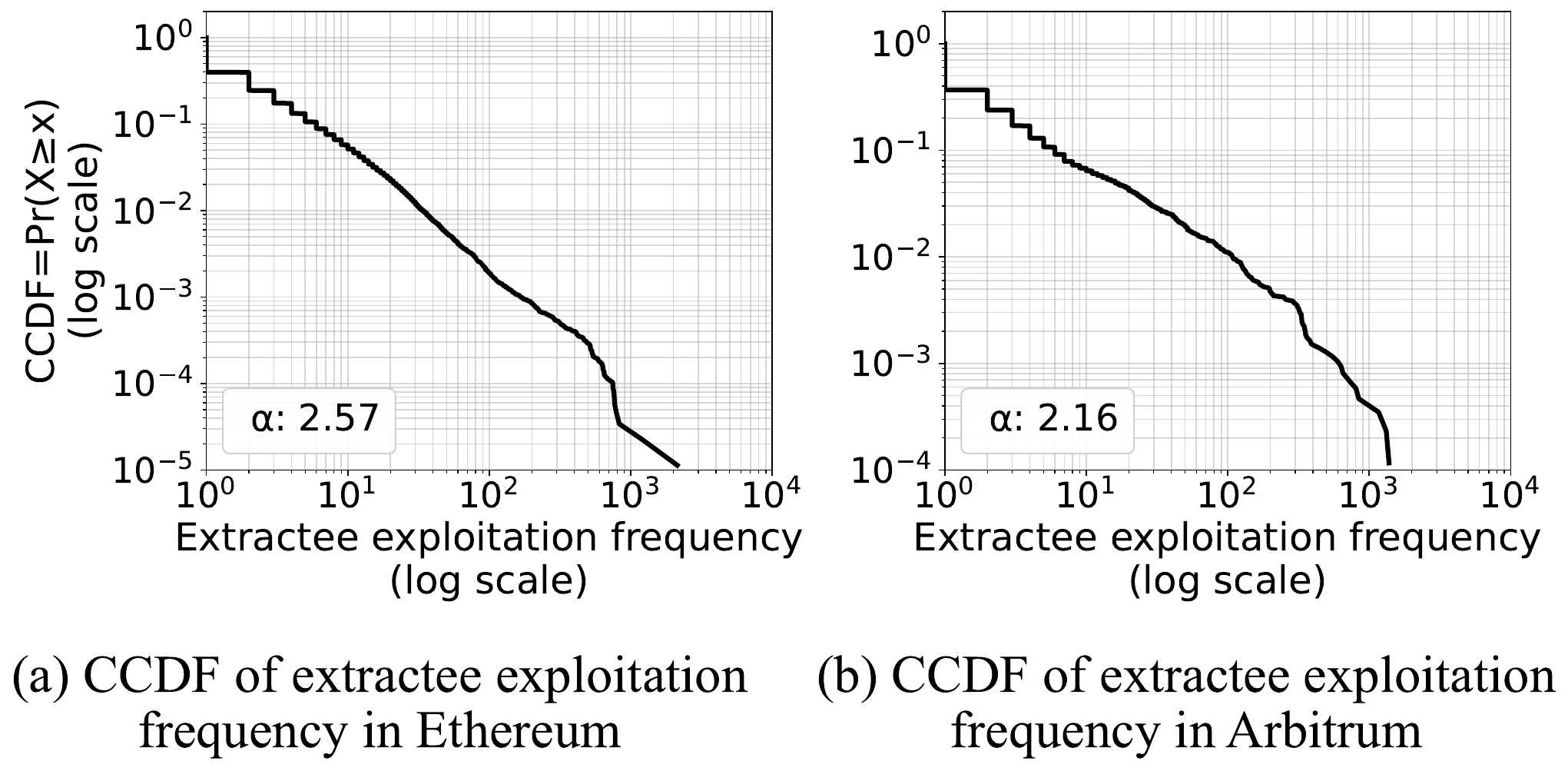}
    \caption{CCDF of [F4] extractee exploitation frequency on Ethereum (a) and Arbitrum (b).}
    \label{fig:extractee-exploitation-frequency}
\end{figure}

\para{[F4] Extractee exploitation frequency.}
\label{sec:extractee-exploitation-frequency}
The CCDF in Figure~\ref{fig:extractee-exploitation-frequency} is again linear on a log-log scale, with fitted exponents of $\alpha = 2.57$ on Ethereum and $\alpha = 2.16$ on Arbitrum.
This heavy tail means that wallets repeatedly losing capital are a common feature of DEX trading rather than an anomaly: a \name sender that distributes funds to many distinct receivers via repeated staged losses cannot be separated by this feature alone from repeated-loss behavior observed in ordinary DEX activity.

\subsection{Inherent Limitations of Threshold-Based Semantic Inspection}
\label{sec:limitations}

The consequence for forensic analysis is that threshold-based semantic inspection has no stable operating point.
A power-law tail is scale-free; namely, there is no characteristic magnitude at which an observation becomes categorically separate from the rest of the distribution.
For the unbounded features, such as extraction volume and frequency-based measures, the distribution has no definitive upper bound, and increasingly large observations remain part of the same empirical tail.
Moreover, for tail exponents in the range observed in our data, these unbounded values can drive the variance to infinity, even though the sample mean and variance can always be computed on any finite dataset.\footnote{Power-law distributions with $\alpha \leq 2$ have no finite mean, while those with $\alpha \leq 3$ have no finite variance~\cite{newman2005power}.}
For the bounded capital-extraction ratio, the analogous phenomenon appears near the boundary; that is, values can approach full extraction arbitrarily closely, and the observed tail indicates that near-boundary losses are not isolated measurement artifacts.

This tail behavior makes any fixed threshold inherently brittle.
A low threshold {could} flag many {false positives}, while a high threshold could allow carefully staged \name transfers to blend into the tail of other MEV activity.
Thus, magnitude alone cannot provide a definitive separation of deliberate DCAT transfers, even when the observed loss is unusually large or repeatedly associated with the same address pair.

This limitation also explains why threshold-based evidence provides {\em plausible deniability} rather than attribution.
Semantic features may still be useful for prioritizing suspicious incidents, but they are insufficient as deterministic indicators.
This motivates the rank-based analysis in the next section, which treats semantic evidence as a triage signal rather than as a binary detection rule.

\section{Rank-Based Analysis and Case Studies}
\label{sec:multivariate-detection}

As shown in previous sections, \name's syntactic mimicry prevents definitive attribution from transaction structure alone, while the power-law characteristics of extraction features make fixed-threshold detection brittle.
Faced with these combined limitations, we use a rank-based approach to prioritize suspicious incidents for manual review.
This section first explains why single-feature ranking is insufficient~(\S\ref{sec:feature-dependencies}), then presents a multivariate ranking analysis method~(\S\ref{sec:trace}), and finally applies the framework to real-world on-chain data through three case studies~(\S\ref{sec:case-study}).

{Note that, as we} rely only on public ledger data, we do not conclusively label {these cases as \name transfers}.
To achieve definitive attribution, this statistical foundation would have to be combined with auxiliary information beyond on-chain data. 
For example, investigators may combine the highest-ranked cases with auxiliary records, such as KYC information or known illicit-address labels, to assess whether a suspicious pattern corresponds to covert value movement. 

\subsection{Rank-Based Approach and Feature Dependencies} 
\label{sec:feature-dependencies}
\begin{figure}[t]
    \centering
    \includegraphics[width=0.5\textwidth]{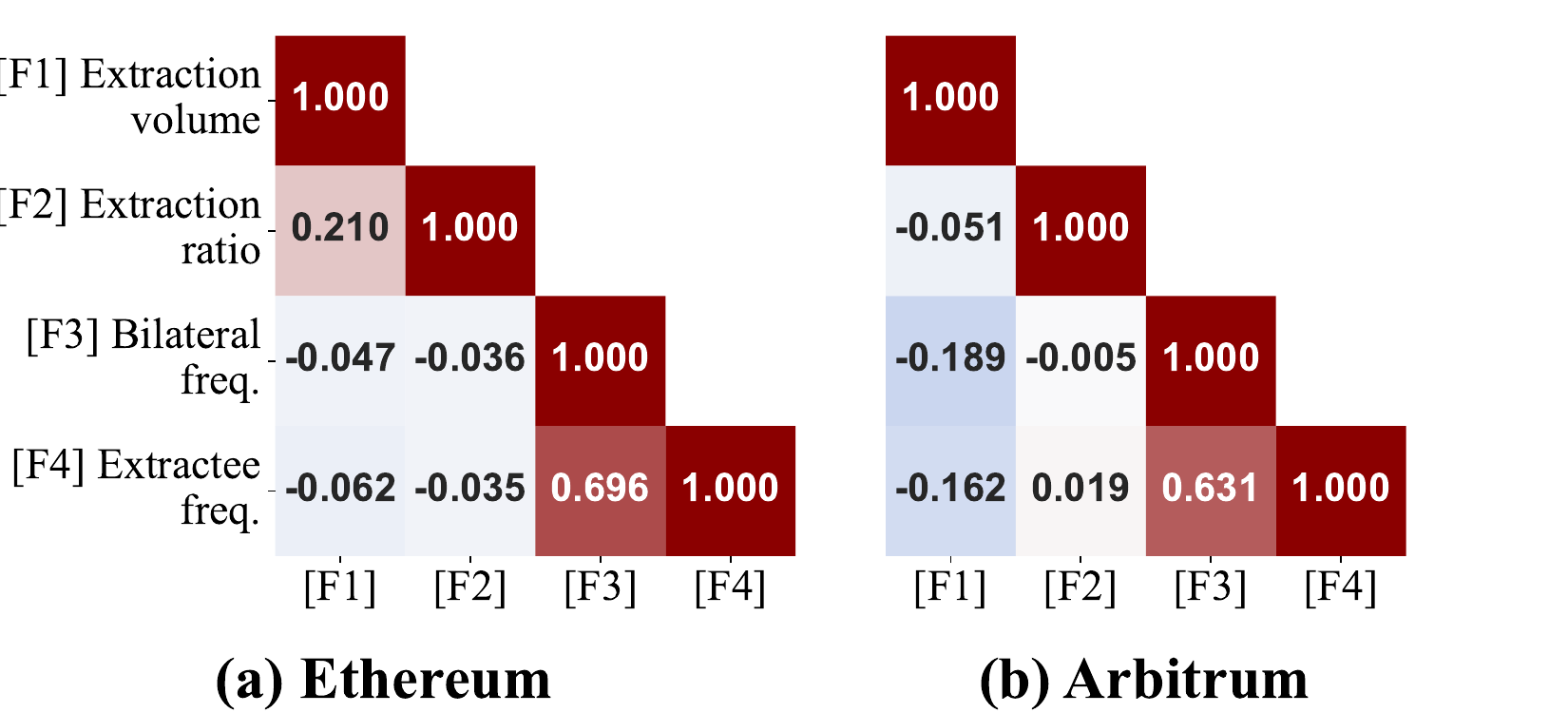}
    \caption{Kendall $\tau$ correlation matrix of the four features on Ethereum (a) and Arbitrum (b).}
    \label{fig:kendall-tau}
\end{figure}

As discussed earlier, the power-law characteristics make it difficult to choose a fixed threshold that {categorically} separates \name incidents from background MEV activity. 
Consequently, investigators would have to manually inspect suspicious cases by starting with the extreme outliers of each feature. 
While capable of surfacing extreme anomalies, this single-feature ranking approach leaves analysts with two practical limitations. 
First, analysts lack a clear method to prioritize among different features when navigating numerous cases.
Second, this isolated view may miss incidents that appear {ordinary} when inspected independently but prove anomalous when two or more features are jointly considered. 

To understand the relationships between these features, we first measure their pairwise dependence using Kendall's $\tau$, as depicted in Figure~\ref{fig:kendall-tau}. 
The results demonstrate that the features are neither fully correlated nor completely independent.
For example, {[F3]} bilateral extraction frequency and {[F4]} extractee exploitation frequency exhibit the strongest dependence on both chains ($\tau = 0.70$ on Ethereum, $\tau = 0.63$ on Arbitrum), as an extractee who is repeatedly victimized tends to be targeted by the same systematic extractor.
The relationship between {[F1]} extraction volume and the remaining three features, however, differs across chains. 
On Ethereum, {[F1]} extraction volume correlates more strongly with the {[F2]} capital-extraction ratio ($\tau = 0.21$) than with the {[F3]-[F4]} frequency features ($\tau \approx -0.05$); conversely, on Arbitrum, this pattern reverses, and volume correlates more strongly with the {[F3]-[F4]} frequency features ($\tau \approx -0.19$). 
This partial dependence structure makes a joint analysis informative; that is, an incident exhibiting mild elevation across all four features simultaneously is far rarer in the observed market than its individual feature values suggest. 
Because most feature pairs are only weakly correlated, such a joint elevation is unlikely to arise naturally. 

{Leveraging this finding}, we focus on a multivariate view that examines the joint footprint of an incident across all four features simultaneously. 
This multivariate approach does not completely replace {per-feature analysis}; rather, it serves as a complementary tool that helps analysts grasp the broader context and prioritize manual review.

\subsection{The Multivariate Ranking Analysis}
\label{sec:trace}
Building on these observations, we implement a statistical multivariate ranking analysis that scores each incident by the \emph{joint} survival probability of its multivariate footprint, by which the incidents are ranked. 
We first model the joint distribution of baseline extractions using a copula, a statistical tool that captures the dependence structure among multiple variables independently of their individual marginal distributions.
{Given the variety of copula families (e.g., Gaussian, Clayton, Gumbel), we carefully select the appropriate one to accurately capture the dataset's underlying dependence structure. }
Accordingly, we fit a $t$-copula ($\nu = 13$) to the Ethereum dataset and a Gaussian copula to the Arbitrum dataset; we defer the detailed justification of these choices to Appendix~\ref{sec:copula}.

{After fitting the appropriate copula, we derive the corresponding survival copula $\hat{C}$ to rank the incidents based on their joint survival probability $p$, where a smaller value indicates a rarer joint profile.}
More formally, for an observed incident with feature values $(x_1, \dots, x_d)$, this joint probability $p$ is calculated using the complementary cumulative distribution function (or survival function):
$$p = \Pr(X_1 > x_1, \dots, X_d > x_d) = \hat{C}(\bar{F}_1(x_1),\dots,\bar{F}_d(x_d)),$$
{where $\bar{F}_i(x_i)$ represents the continuous marginal complementary cumulative distribution function evaluated at the observed value $x_i$ of the $i$-th feature.}
Consequently, an analyst can prioritize investigations by ranking cases in ascending order of $p$, focusing manual review efforts on the most statistically improbable events.
\begin{figure}[t]
    \centering
    \includegraphics[width=0.45\textwidth]{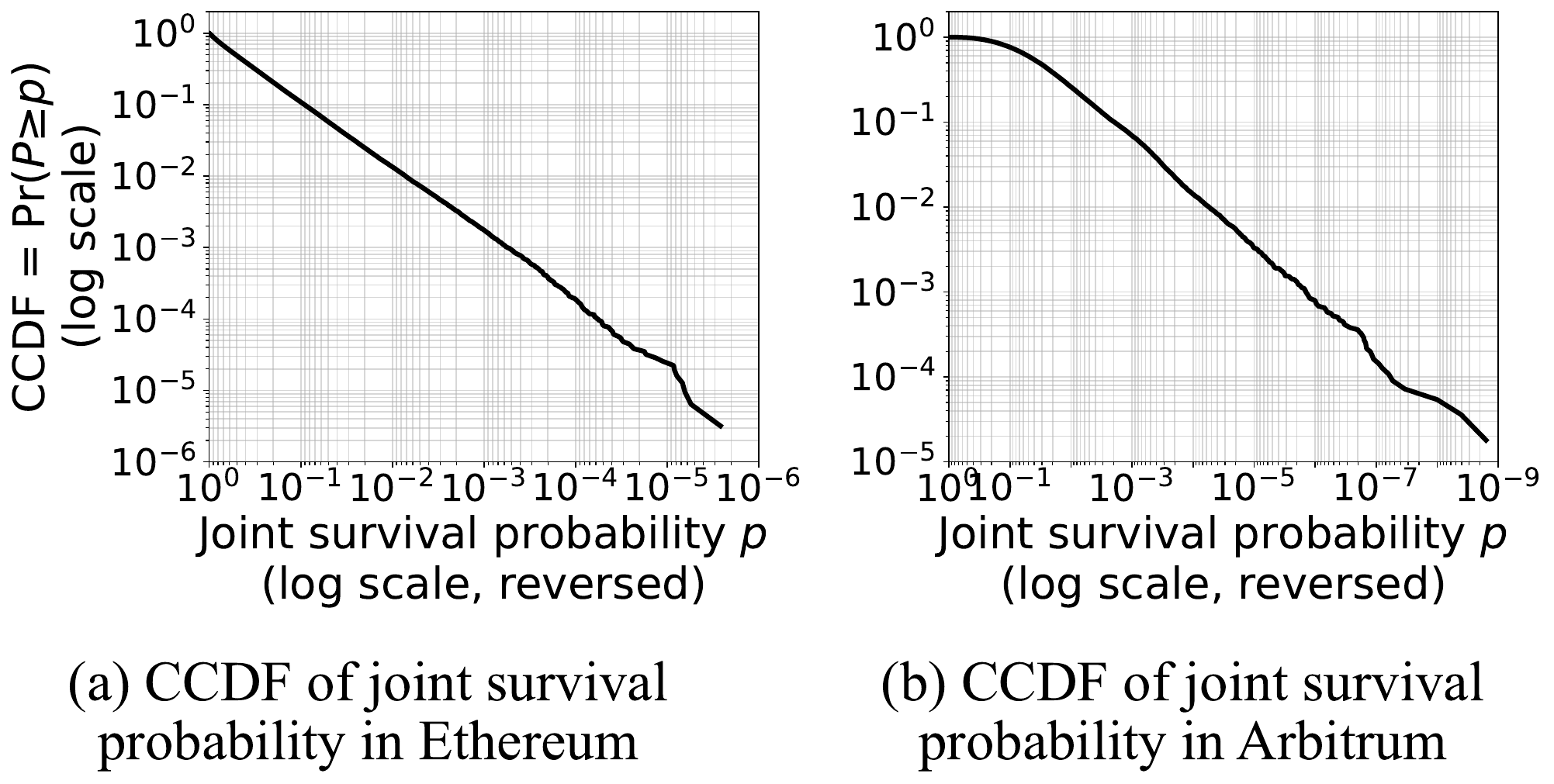}
    \caption{CCDF of the joint survival probability $p$ calculated using a survival copula on Ethereum (a) and Arbitrum (b). 
    Smaller $p$ indicates a rarer case.}
    \label{fig:joint-proba}
\end{figure}

\para{Results.}
We apply this analysis to our full dataset of 313{,}219 Ethereum sandwich incidents and 55{,}482 Arbitrum arbitrage incidents; revisit \S\ref{sec:empirical-baselines} for details.
Figure~\ref{fig:joint-proba} shows the CCDF of the joint survival probability $p$ for both chains.
Notably, among the top 500 lowest-$p$ incidents on Ethereum with the lowest $p$ scores, 52\% exhibit no individual feature above the 90th percentile; 
{this means that they would likely be missed when observing single features alone, whereas they are surfaced by examining their joint multivariate footprint.}
On Arbitrum, the corresponding fraction is zero, indicating that the multivariate approach primarily provides a unified ranking of suspiciousness rather than uncovering entirely new candidates. 

Notably, the log-log CCDF in Figure~\ref{fig:joint-proba} again exhibits a straight line, suggesting that the joint survival probability may also exhibit a power-law characteristic. 
As discussed in the previous section~(\S\ref{sec:limitations}), this implies that it is exceedingly challenging to establish a {fixed} anomaly-detection threshold, making it difficult to deterministically classify extreme outliers as anomalous purely based on their $p$-values. 
In other words, the joint survival probability $p$ serves as a continuous ranking metric to prioritize investigations rather than as a strict binary classifier for anomalies. 

\subsection{Real-world Case Studies}
\label{sec:case-study}

We now examine three example cases drawn from the extreme tail of our multivariate analysis.
These cases are not intended to establish {definitive} attribution.
Rather, they illustrate how the proposed ranking framework can prioritize incidents that merit deeper investigation using only public on-chain data.
Definitive attribution would require auxiliary information, such as case records, exchange records, or known illicit-address labels.

Table~\ref{tab:case-study-comparison} summarizes the three case studies and the distinct forensic signal each one illustrates.
Case~1 captures systematic repetition in Arbitrum arbitrage, Case~2 captures extreme extraction volume in Ethereum sandwiches, and Case~3 captures a jointly rare Ethereum footprint that is not extreme along any single feature.

\begin{table*}[t]
\centering
\caption{Three suspicious MEV cases surfaced by our rank-based analysis on Ethereum and Arbitrum.}
\label{tab:case-study-comparison}
\scriptsize
\begin{tabular}{@{}lllrll@{}}
\toprule
No. & Month/Year & Chain, MEV type & Inferred transferred value & Main signal & Core investigative clue \\
\midrule
Case 1 & Mar.-Apr. 2025 & Arbitrum, arbitrage & \$43.9K & Repeated extraction & One extractee, 27 extractors, large fan-out pattern \\
Case 2 & Mar. 2025 & Ethereum, sandwich & \$502.8K & Extreme volume & High-ranked incidents share timing, extractor, and funding source \\
Case 3 & May.-Aug. 2025 & Ethereum, sandwich & \$2.8K & Jointly rare footprint & Not extreme per-feature, but surfaced by multivariate ranking \\
\bottomrule
\end{tabular}
\end{table*}

\subsubsection{\bf\em Case 1: Systematic extraction on Arbitrum}
\label{sec:case-1} 

\begin{figure}[t]
    \centering
    \includegraphics[width=0.45\textwidth]{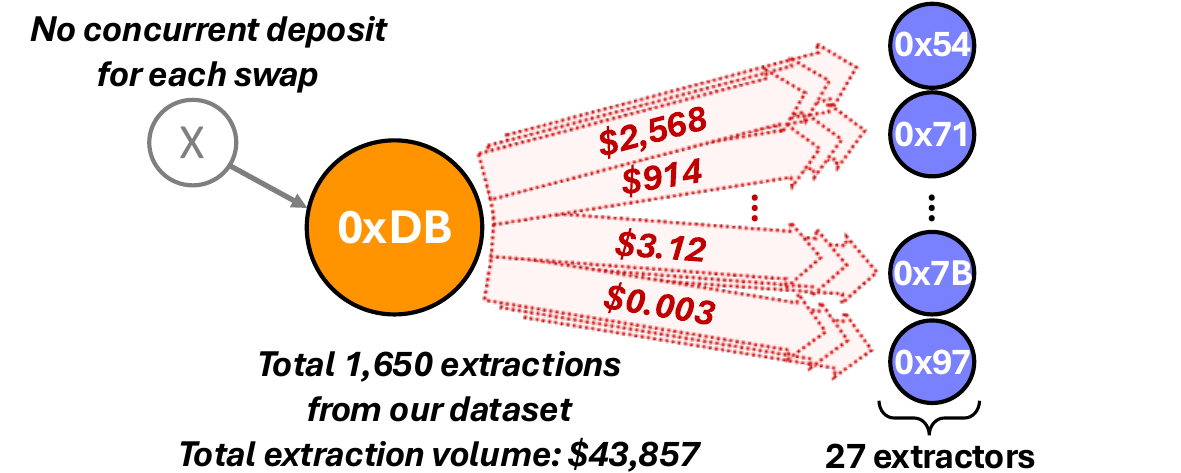}
    \caption{Case 1. On Arbitrum, extractee \addr{0xDB} was arbitraged 1{,}650 times over two months.
    In total, 27 extractors captured \$43{,}857, with the top extractor securing \$6{,}250. 
    }
    \label{fig:case-2}
\end{figure}

We first present a long-running Arbitrum campaign, shown in Figure~\ref{fig:case-2}, that is visible to both per-feature and multivariate ranking.  
{
Within our observed dataset (March 1 to April 20, 2025), the extractee \addr{0xDB}\footnote{\url{https://arbiscan.io/address/0xdb633ffd9d74afa6782eb753349bb8b4e2b51cc2}} was targeted 1{,}650 times by 27 distinct extractors, yielding a total extraction volume of \$43,857.
The most profitable extractor (\addr{0x54}) accumulated roughly \$6{,}251, and the single largest extraction against this extractee\footnote{\url{https://arbiscan.io/tx/0xf07e8957eab98954ebfcae4f6529129c2a0c182dbbf519402072fbaf807ef836}} netted \$2{,}569 at a 0.30 loss ratio. 

Several features place this campaign near the extreme tail.
The largest single extraction against this address (a profit of \$2{,}569) ranks 14th by volume across our Arbitrum dataset; the highest bilateral frequency from a single extractor (161 times) ranks 18th; and the extractee's overall exploitation frequency (1{,}650 times) ranks first among all Arbitrum extractees. 
The multivariate framework also ranks the headline arbitrage incident involving this extractee as the second most anomalous event on Arbitrum. 
}

{
Manual inspection reveals that \addr{0xDB} also executed 
32{,}133 WETH-to-ARB swaps on a single Uniswap pool (\addr{0xBF}) at intervals of roughly two minutes; in total, this converted 4{,}591.18~WETH ($\approx\$7.8M$) into 25{,}167{,}463.30~ARB ($\approx\$7.5M$) between March 12 and May 16, 2025. 
Two further observations highlight the anomalous nature of this address. 
First, while the address might initially appear to be a service provider given that the acquired ARB was forwarded to 8{,}232 distinct recipient addresses in a fan-out pattern, the WETH used for the swaps originated entirely from the extractee itself. 
If this entity were acting as a legitimate service provider, it should have received funds from external sources concurrently, thereby exhibiting a simultaneous fan-in and fan-out structure. 
We found no public information linking this address to a known service, leaving it as an unexplained, unilateral token distributor. 
Second, the address ceased all on-chain activity in May 2025 immediately after its WETH balance was exhausted. 
}
Under a single-operator interpretation, this cluster would correspond to a covert transfer of \$43,857 during our observation window, possibly alongside a separate swap-and-fan-out flow.

\subsubsection{\bf\em Case 2: A cluster of large extractions on Ethereum}
\label{sec:case-2} 

\begin{figure}[t]
    \centering
    \includegraphics[width=0.45\textwidth]{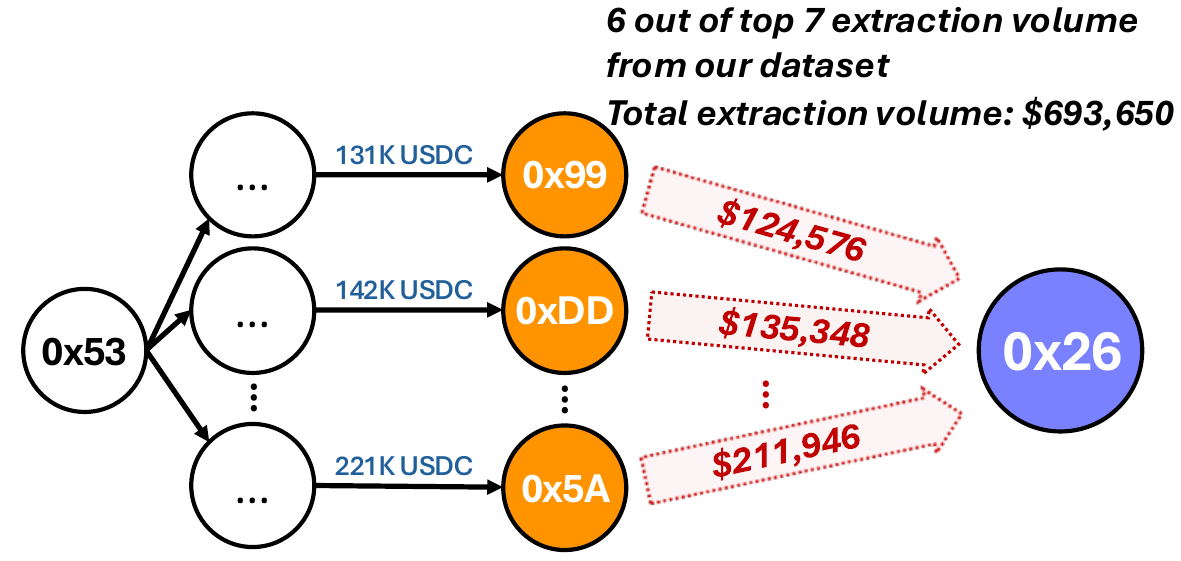} 
    \caption{Case 2. Six seemingly independent sandwich extractions occurred within five minutes. 
    The six extractees were funded by the same source, interacted with the same DEX, and extracted by the same extractor, yielding a total extracted revenue of \$693{,}650.} 
    \label{fig:case-1} 
\end{figure}

We next examine {six} high-value Ethereum sandwich extractions, shown in Figure~\ref{fig:case-1}. 
On March 12, 2025, around 09:00 UTC, the same extractor (\addr{0x26}) sandwiched {six} extractees within a four-minute window. 
The extractor captured approximately \$54{,}446 from \addr{0x5a}\footnote{\url{https://etherscan.io/tx/0x937d424f43eeb36ec91651e1c9a48a8d7c9f8a457e09375b9e8284743e7ec5e4}},
\$82{,}277 from \addr{0x2b}\footnote{\url{https://etherscan.io/tx/0xc7bb7130407ac0d5d94b8a9b682ac2c32a9d4292454eca39b35220d0321c76f9}},
\$85{,}057 from \addr{0x5D}\footnote{\url{https://etherscan.io/tx/0x14b910b4044eb0c64ce080c02e836eec8302f0fcfe3ffc5c757373bfba313845}},
\$124{,}576 from \addr{0x99}\footnote{\url{https://etherscan.io/tx/0x636ca7a5f63a6698c82e860e610363e75d97184403fa06f9907243a84e25b68a}},
\$135{,}348 from \addr{0xDD}\footnote{\url{https://etherscan.io/tx/0xb403f921671c5f1494153f0b9543a0650ff1212a4b557f2ea53455107c94215c}}, and \$211{,}946 from \addr{0x5A}\footnote{\url{https://etherscan.io/tx/0xee9fcd2b9996e96b642cb4cda47fc140f98fdaf07ee02657743d4bfcc4670106}}, yielding \$693{,}650 in total extracted revenue.
The captured value was subsequently distributed among the extractor, the block builder, and the proposers.\footnote{The extractor received \$12{,}463, \$10{,}684, \$11{,}287, \$13{,}927, \$13{,}444, and \$20{,}090; the builder received \$30{,}152, \$42{,}732, \$52{,}910, \$87{,}173, \$76{,}160, and \$131{,}730; and the proposers received \$11{,}832, \$28{,}861, \$20{,}861, \$34{,}249, \$34{,}972, and \$60{,}125 for the extractions of \addr{0x5a}, \addr{0x2b}, \addr{0x5D}, \addr{0x99}, \addr{0xDD}, and \addr{0x5A}, respectively.}

{
{These six extractions rank first through fifth, and seventh} by volume across our entire Ethereum dataset, ensuring that single-feature ranking flags them immediately. 
The multivariate framework ranks these same incidents {86th, 157th, 200th, 233rd, 264th, and 349th, respectively.}
This illustrates a complementary role of the two views: when one feature is already extreme, a single-feature ranking is the primary signal, while the multivariate score provides corroborating context.
}

{
Further manual inspection of public traces suggest that the six extractees are not independent. 
Their funds trace back through separate intermediate paths to a common source address (\addr{0x53}), which routed parallel transfers to each extractee.
Two additional observations sharpen this picture. 
First, the intermediate addresses on each funding path executed swaps with conventional slippage and incurred minimal losses, indicating that the operator is capable of configuring safe slippage tolerances. 
Second, all six extractees ceased on-chain activity shortly after their respective extractions. 
Note that this is a rediscovery of a previously reported large extraction case~\cite{tim2025Money,cointelegraph2025crypto,cryptoalchemy2025defi}, but our case study provides a more comprehensive and systematic analysis of this highly suspicious group of extractions.}

If a single entity controlled the six extractees, the extractor, and the builder, this cluster would correspond to a net asset movement of \$502{,}752 at a cost of \$190{,}900 paid to the proposers. 
A benign explanation remains possible, but the timing, common funding source, shared extractor, and post-extraction inactivity make this cluster a high-priority candidate for manual review.

\subsubsection{\bf\em Case 3: Inconspicuous extractions on Ethereum}
\label{sec:case-3} 

\begin{figure}[t]
    \centering
    \includegraphics[width=0.45\textwidth]{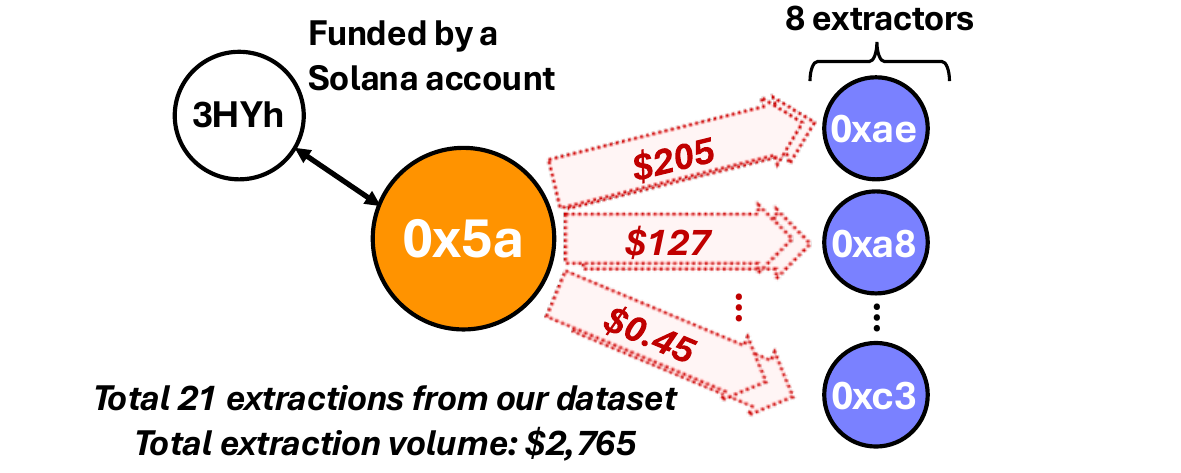}
    \caption{Case 3. Extractee \addr{0x5a} was sandwiched 21 times by a total of eight extractors, resulting in a total extraction volume of \$2{,}765. 
    Although each individual feature remains within a normal range, the joint probability of this footprint is highly improbable.} 
    \label{fig:case-3}
\end{figure}

Finally, we present a case, depicted in Figure~\ref{fig:case-3}, that demonstrates the specific value of multivariate ranking.
{
The multivariate framework ranks a sandwich incident involving extractee \addr{0x5a}\footnote{\url{https://etherscan.io/address/0x5a7aacccca80cb06d65100c91cd6da1c4d87590c}} as the 141st most anomalous incident on Ethereum.
Across our dataset, \addr{0x5a} was sandwiched 21 times by eight distinct extractors, losing a total of \$2{,}765 at an average extraction ratio of 0.02. 
The most active extractor (\addr{0xae}) accounted for 12 of these incidents and \$1{,}269 in extracted value. 
}

{
No single feature is individually extreme.
The largest extraction volume is \$374 (ranked 2{,}257th), the highest loss ratio is 0.07 (ranked 43{,}725th), the bilateral frequency is 12 (ranked 2{,}914th), and the extractee exploitation frequency is 21 (ranked 1{,}942nd). 
Thus, a single-feature review would need to inspect more than 1,900 higher-ranked Ethereum incidents before reaching this extractee.
The multivariate framework instead places the case 141st overall, making it much more likely to receive manual attention.
}

{
Further manual inspection reveals a distinctive cross-chain pattern.
The extractee was initially funded by a Solana address (\addr{3HYh}) via a Solana-to-Ethereum bridge. 
After a period of activity, \addr{0x5a} bridged its remaining balance back to the same Solana address and ceased all on-chain operations. 
If both the extractee and its extractors were controlled by a single \name adversary, this campaign would constitute a covert transfer of \$2{,}765 from Solana to Ethereum, executed without any observable on-chain link between the sender and the receiver. 
}
\section{Discussions}
\label{sec:discussion}

We discuss two questions related to \name: 
Is \name applicable to other economic loss events beyond sandwich and arbitrage extractions?~{(\S\ref{sec:other-loss-events})}
Can \name be possibly eradicated?~{(\S\ref{sec:eradicating})}

\subsection{Other Possible Economic Loss Events for \name}
\label{sec:other-loss-events}
Our experiments focus on MEV extractions, but \name only requires an on-chain event in which one party appears to lose value and another party can plausibly capture it as ordinary market profit.
Several other DeFi mechanisms may therefore provide similar camouflage.

\begin{packeditemize}
\item {\em Liquidation.}
Decentralized lending protocols use liquidation to resolve undercollateralized positions~\cite{ferreira2024rolling,qin2021empirical}. 
When the value of a borrower's collateral falls below a protocol-defined safety threshold, the smart contract liquidates the position by auctioning the collateral at a discounted price to recover the outstanding debt. 
A \name sender could deliberately create an undercollateralized position and allow it to be liquidated, while the receiver wins the liquidation and captures the discounted collateral.

\item {\em Rug pull.}
A rug pull commonly involves a project operator creating a trading pair, attracting swaps into a low-value or worthless token, and then withdrawing the valuable liquidity~\cite{cernera2023token}.
\name can reuse this pattern by having the receiver deploy the trading pair and withdraw liquidity after the sender intentionally swaps the target asset into the pool.
The resulting loss appears as participation in a fraudulent or failed market rather than as a direct transfer.
\end{packeditemize}

These examples are not exhaustive.
They suggest that \name is best understood as a general way to hide value movement inside ecosystem-native loss events, not as a technique specific to the two MEV settings evaluated in this paper.

\subsection{Eradicating \name}
\label{sec:eradicating}

Completely eradicating \name appears unlikely, unfortunately, without removing the loss events that provide its cover.
Prevalent practices such as sandwich extraction and arbitrage, however, are deeply integrated into the structural design and economic incentives of many blockchain ecosystems~\cite{PBS-ethereum,qin2022quantifying}, rendering their complete removal impractical.
As long as MEV activities persist in the ecosystem and are considered normal, \name can continue to operate by utilizing these events as cover for unobservable transfers.

There have been efforts to mitigate MEV extraction through technical means, such as fair ordering protocols~\cite{kelkar2023themis, mu2024separation}.
However, no technical mitigation has yet demonstrated the near-complete elimination of MEV extraction.
Worse yet, recent work shows that technical mitigations against MEV extraction may be bypassed in some fair-ordering protocols~\cite{park2025frontrunning}.
Overall, we argue that \name is more realistically addressed through forensic triage, ecosystem monitoring, and auxiliary investigation than through complete eradication.

\section{Conclusion}
\label{sec:conclusion}

This paper studied whether staged economic loss events in blockchains can covertly move assets while making the transfer itself unobservable.
To our knowledge, this is the first in-depth study showing that this stronger form of untraceability can be instantiated on representative layer-1 and layer-2 blockchains through \name, which blends transfers into background MEV activity.
We further showed that semantic inspection has limited separating power due to heavy-tailed MEV behavior, but that multi-feature outlier ranking can still support forensic triage.
By extending blockchain untraceability from unlinkability to unobservability, this work adds a new analytical lens to blockchain forensics for reasoning about future covert value-movement cases.

\subsection*{Ethical Considerations}
\label{sec:ethics}

This work poses minimal ethical risk.
All end-to-end \name experiments were conducted on public testnets, specifically the Ethereum Hoodi testnet and the Arbitrum Sepolia testnet.
Even in these testnet environments, our experiments used only wallets, tokens, and liquidity pools controlled by the authors.
Thus, the experiments did not interact with or affect other users, assets, or applications on the testnets.
Our real-world analyses rely only on public on-chain data and are used for aggregate measurement and forensic triage, not for definitive attribution of specific entities.

\bibliographystyle{IEEEtran}
\bibliography{main}

@inproceedings{sasson2014zerocash,
  title={Zerocash: Decentralized anonymous payments from bitcoin},
  author={Sasson, Eli Ben and Chiesa, Alessandro and Garman, Christina and Green, Matthew and Miers, Ian and Tromer, Eran and Virza, Madars},
  booktitle={2014 IEEE symposium on security and privacy},
  pages={459--474},
  year={2014},
  organization={IEEE}
}

@inproceedings{wu2021towards,
  title={Towards understanding and demystifying bitcoin mixing services},
  author={Wu, Lei and Hu, Yufeng and Zhou, Yajin and Wang, Haoyu and Luo, Xiapu and Wang, Zhi and Zhang, Fan and Ren, Kui},
  booktitle={Proceedings of the Web Conference 2021},
  pages={33--44},
  year={2021}
}

@misc{ustreasury22,
    title = {U.S. Treasury Sanctions Notorious Virtual Currency Mixer Tornado Cash},
    author = {U.S. Department of the Treasury},
    howpublished = {\url{https://home.treasury.gov/news/press-releases/jy0916}},
    year = {2022},
}

@misc{van2013cryptonote,
    title={CryptoNote v 2.0},
    author={Van Saberhagen, Nicolas},
    year={2013},
    howpublished = {\url{https://web.getmonero.org/resources/research-lab/pubs/cryptonote-whitepaper.pdf}},
}

@inproceedings{lubbertsen2025ghost,
  title={Ghost Clusters: Evaluating Attribution of Illicit Services through Cryptocurrency Tracing},
  author={Lubbertsen, Kelvin and van Eeten, Michel and van Wegberg, Rolf},
  booktitle={34th USENIX Security Symposium (USENIX Security 25)},
  pages={1357--1374},
  year={2025}
}

@inproceedings{huang2018tracking,
  title={Tracking ransomware end-to-end},
  author={Huang, Danny Yuxing and Aliapoulios, Maxwell Matthaios and Li, Vector Guo and Invernizzi, Luca and Bursztein, Elie and McRoberts, Kylie and Levin, Jonathan and Levchenko, Kirill and Snoeren, Alex C and McCoy, Damon},
  booktitle={2018 IEEE Symposium on Security and Privacy (SP)},
  pages={618--631},
  year={2018},
  organization={IEEE}
}

@inproceedings{lin2024denseflow,
  title={DenseFlow: Spotting cryptocurrency money laundering in ethereum transaction graphs},
  author={Lin, Dan and Wu, Jiajing and Yu, Yunmei and Fu, Qishuang and Zheng, Zibin and Yang, Changlin},
  booktitle={Proceedings of the ACM Web Conference 2024},
  pages={4429--4438},
  year={2024}
}

@article{wu2023tracer,
  title={TRacer: Scalable graph-based transaction tracing for account-based blockchain trading systems},
  author={Wu, Zhiying and Liu, Jieli and Wu, Jiajing and Zheng, Zibin and Chen, Ting},
  journal={IEEE Transactions on Information Forensics and Security},
  volume={18},
  pages={2609--2621},
  year={2023},
  publisher={IEEE}
}

@inproceedings{victor2020address,
  title={Address clustering heuristics for Ethereum},
  author={Victor, Friedhelm},
  booktitle={International conference on financial cryptography and data security},
  pages={617--633},
  year={2020},
  organization={Springer}
}

@inproceedings{meiklejohn2013fistful,
  title={A fistful of bitcoins: characterizing payments among men with no names},
  author={Meiklejohn, Sarah and Pomarole, Marjori and Jordan, Grant and Levchenko, Kirill and McCoy, Damon and Voelker, Geoffrey M and Savage, Stefan},
  booktitle={Proceedings of the 2013 conference on Internet measurement conference},
  pages={127--140},
  year={2013}
}

@inproceedings{androulaki2013evaluating,
  title={Evaluating user privacy in bitcoin},
  author={Androulaki, Elli and Karame, Ghassan O and Roeschlin, Marc and Scherer, Tobias and Capkun, Srdjan},
  booktitle={International conference on financial cryptography and data security},
  pages={34--51},
  year={2013},
  organization={Springer}
}

@incollection{reid2012analysis,
  title={An analysis of anonymity in the bitcoin system},
  author={Reid, Fergal and Harrigan, Martin},
  booktitle={Security and privacy in social networks},
  pages={197--223},
  year={2012},
  publisher={Springer}
}

@inproceedings{ron2013quantitative,
  title={Quantitative analysis of the full bitcoin transaction graph},
  author={Ron, Dorit and Shamir, Adi},
  booktitle={International conference on financial cryptography and data security},
  pages={6--24},
  year={2013},
  organization={Springer}
}

@inproceedings{liao2016behind,
  title={Behind closed doors: measurement and analysis of CryptoLocker ransoms in Bitcoin},
  author={Liao, Kevin and Zhao, Ziming and Doup{\'e}, Adam and Ahn, Gail-Joon},
  booktitle={2016 APWG symposium on electronic crime research (eCrime)},
  pages={1--13},
  year={2016},
  organization={IEEE}
}

@inproceedings{spagnuolo2014bitiodine,
  title={Bitiodine: Extracting intelligence from the bitcoin network},
  author={Spagnuolo, Michele and Maggi, Federico and Zanero, Stefano},
  booktitle={International conference on financial cryptography and data security},
  pages={457--468},
  year={2014},
  organization={Springer}
}

@inproceedings{gomez2022watch,
  title={Watch your back: Identifying cybercrime financial relationships in bitcoin through back-and-forth exploration},
  author={Gomez, Gibran and Moreno-Sanchez, Pedro and Caballero, Juan},
  booktitle={Proceedings of the 2022 ACM SIGSAC conference on computer and communications security},
  pages={1291--1305},
  year={2022}
}

@inproceedings{kappos2022peel,
  title={How to peel a million: Validating and expanding bitcoin clusters},
  author={Kappos, George and Yousaf, Haaroon and St{\"u}tz, Rainer and Rollet, Sofia and Haslhofer, Bernhard and Meiklejohn, Sarah},
  booktitle={31st USENIX security symposium (USENIX security 22)},
  pages={2207--2223},
  year={2022}
}

@inproceedings{pfitzmann2001anonymity,
  title={Anonymity, unobservability, and pseudonymity—a proposal for terminology},
  author={Pfitzmann, Andreas and K{\"o}hntopp, Marit},
  booktitle={Designing Privacy Enhancing Technologies: International Workshop on Design Issues in Anonymity and Unobservability Berkeley, CA, USA, July 25--26, 2000 Proceedings},
  pages={1--9},
  year={2001},
  organization={Springer}
}

@inproceedings{miers2013zerocoin,
  title={Zerocoin: Anonymous distributed e-cash from bitcoin},
  author={Miers, Ian and Garman, Christina and Green, Matthew and Rubin, Aviel D},
  booktitle={2013 IEEE symposium on security and privacy},
  pages={397--411},
  year={2013},
  organization={IEEE}
}

@inproceedings{qin2022quantifying,
  title={Quantifying blockchain extractable value: How dark is the forest?},
  author={Qin, Kaihua and Zhou, Liyi and Gervais, Arthur},
  booktitle={2022 IEEE Symposium on Security and Privacy (SP)},
  pages={198--214},
  year={2022},
  organization={IEEE}
}

@inproceedings{daian2020flash,
  title={Flash boys 2.0: Frontrunning in decentralized exchanges, miner extractable value, and consensus instability},
  author={Daian, Philip and Goldfeder, Steven and Kell, Tyler and Li, Yunqi and Zhao, Xueyuan and Bentov, Iddo and Breidenbach, Lorenz and Juels, Ari},
  booktitle={2020 IEEE symposium on security and privacy (SP)},
  pages={910--927},
  year={2020},
  organization={IEEE}
}

@inproceedings{wang2023zero,
  title={On how zero-knowledge proof blockchain mixers improve, and worsen user privacy},
  author={Wang, Zhipeng and Chaliasos, Stefanos and Qin, Kaihua and Zhou, Liyi and Gao, Lifeng and Berrang, Pascal and Livshits, Benjamin and Gervais, Arthur},
  booktitle={Proceedings of the ACM Web Conference 2023},
  pages={2022--2032},
  year={2023}
}

@article{wu2023toward,
  title={Toward understanding asset flows in crypto money laundering through the lenses of Ethereum heists},
  author={Wu, Jiajing and Lin, Dan and Fu, Qishuang and Yang, Shuo and Chen, Ting and Zheng, Zibin and Song, Bowen},
  journal={IEEE Transactions on Information Forensics and Security},
  volume={19},
  pages={1994--2009},
  year={2023},
  publisher={IEEE}
}

@misc{rolling-github,
    title={Github of Rolling in the Shadows},
    author={Ferreira Torres, Christof and Mamuti, Albin and Weintraub, Ben and Nita-Rotaru, Cristina and Shinde, Shweta},
    howpublished={\url{https://github.com/christoftorres/Rolling-in-the-Shadows}},
    year={2024},
    note={Accessed: 2025-11}
}

@misc{mev-inspect-github,
    title={Github repository of mev-inspect-py},
    author={Flashbots},
    howpublished={\url{https://github.com/flashbots/mev-inspect-py}},
    note={Accessed: 2025-11}
}

@misc{lighthouse-github,
    title={Github repository of lighthouse},
    author={Sigma Prime},
    howpublished={\url{https://github.com/sigp/lighthouse}},
    note={Accessed: 2025-11}
}

@misc{reth-github,
    title={Github repository of reth},
    author={Paradigm},
    howpublished={\url{https://github.com/paradigmxyz/reth}},
    note={Accessed: 2025-11}
}

@misc{mev-boost-github,
    title={Github repository of mev-boost},
    author={Flashbots},
    howpublished={\url{https://github.com/flashbots/mev-boost}},
    note={Accessed: 2025-11}
}

@misc{flashbot-relay-hoodi,
    title={Flashbots Boost Relay - Hoodi},
    author={Flashbots},
    howpublished={\url{https://boost-relay-hoodi.flashbots.net/}},
    note={Accessed: 2025-11}
}

@misc{rbuilder-github,
    title={Github repository of rbuilder},
    author={Flashbots},
    howpublished={\url{https://github.com/flashbots/rbuilder}},
    note={Accessed: 2025-11}
}

@misc{uniswap-github,
    title={Github repository of Uniswap v2-core},
    author={Uniswap},
    howpublished={\url{https://github.com/Uniswap/v2-core}},
    note={Accessed: 2025-11}
}

@misc{coingecko,
    title={CoinGecko},
    author={CoinGecko},
    howpublished={\url{https://www.coingecko.com/}},
    note={Accessed: 2025-11}
}

@article{clauset2009power,
  title={Power-law distributions in empirical data},
  author={Clauset, Aaron and Shalizi, Cosma Rohilla and Newman, Mark EJ},
  journal={SIAM review},
  volume={51},
  number={4},
  pages={661--703},
  year={2009},
  publisher={SIAM}
}

@article{newman2005power,
  title={Power laws, Pareto distributions and Zipf's law},
  author={Newman, Mark EJ},
  journal={Contemporary physics},
  volume={46},
  number={5},
  pages={323--351},
  year={2005},
  publisher={Taylor \& Francis}
}

@inproceedings{moser2013inquiry,
  title={An inquiry into money laundering tools in the Bitcoin ecosystem},
  author={M{\"o}ser, Malte and B{\"o}hme, Rainer and Breuker, Dominic},
  booktitle={2013 APWG eCrime researchers summit},
  pages={1--14},
  year={2013},
  organization={Ieee}
}

@inproceedings{deuber2021coinjoin,
  title={CoinJoin in the wild: An empirical analysis in dash},
  author={Deuber, Dominic and Schr{\"o}der, Dominique},
  booktitle={European Symposium on Research in Computer Security},
  pages={461--480},
  year={2021},
  organization={Springer}
}

@article{alonso2020zero,
  title={Zero to monero},
  author={Alonso, Kurt M and others},
  journal={Zero to monero},
  year={2020}
}

@article{pertsev2019tornado,
  title={Tornado cash privacy solution version 1.4},
  author={Pertsev, Alexey and Semenov, Roman and Storm, Roman},
  journal={Tornado cash privacy solution version},
  volume={1},
  number={6},
  year={2019}
}

@inproceedings{danezis2013pinocchio,
  title={Pinocchio coin: building zerocoin from a succinct pairing-based proof system},
  author={Danezis, George and Fournet, Cedric and Kohlweiss, Markulf and Parno, Bryan},
  booktitle={Proceedings of the First ACM workshop on Language support for privacy-enhancing technologies},
  pages={27--30},
  year={2013}
}

@inproceedings{garman2014rational,
  title={Rational zero: Economic security for zerocoin with everlasting anonymity},
  author={Garman, Christina and Green, Matthew and Miers, Ian and Rubin, Aviel D},
  booktitle={International Conference on Financial Cryptography and Data Security},
  pages={140--155},
  year={2014},
  organization={Springer}
}

@inproceedings{ruffing2014coinshuffle,
  title={Coinshuffle: Practical decentralized coin mixing for bitcoin},
  author={Ruffing, Tim and Moreno-Sanchez, Pedro and Kate, Aniket},
  booktitle={European symposium on research in computer security},
  pages={345--364},
  year={2014},
  organization={Springer}
}

@inproceedings{bonneau2014mixcoin,
  title={Mixcoin: Anonymity for bitcoin with accountable mixes},
  author={Bonneau, Joseph and Narayanan, Arvind and Miller, Andrew and Clark, Jeremy and Kroll, Joshua A and Felten, Edward W},
  booktitle={International conference on financial cryptography and data security},
  pages={486--504},
  year={2014},
  organization={Springer}
}

@inproceedings{valenta2015blindcoin,
  title={Blindcoin: Blinded, accountable mixes for bitcoin},
  author={Valenta, Luke and Rowan, Brendan},
  booktitle={International Conference on Financial Cryptography and Data Security},
  pages={112--126},
  year={2015},
  organization={Springer}
}

@article{turner2018bitcoin,
  title={Bitcoin transactions: a digital discovery of illicit activity on the blockchain},
  author={Turner, Adam and Irwin, Angela Samantha Maitland},
  journal={Journal of Financial Crime},
  volume={25},
  number={1},
  pages={109--130},
  year={2018},
  publisher={Emerald Publishing Limited}
}

@article{farrugia2020detection,
  title={Detection of illicit accounts over the Ethereum blockchain},
  author={Farrugia, Steven and Ellul, Joshua and Azzopardi, George},
  journal={Expert systems with applications},
  volume={150},
  pages={113318},
  year={2020},
  publisher={Elsevier}
}

@misc{uniswap-share,
    title = {Market Share of Decentralized Crypto Exchanges, by Trading Volume},
    author = {Shaun Paul Lee},
    howpublished = {\url{https://www.coingecko.com/research/publications/decentralized-crypto-exchanges-market-share/}},
    year= {2026},
    note= {Accessed: 2026-04}
}

@misc{uniswap-stats,
    key= {Uniswap MAU in 2025},
    title = {Uniswap monthly active users},
    author = {TokenTerminal}, 
    howpublished = {\url{https://tokenterminal.com/explorer/projects/uniswap/metrics/user-mau/}},
    year= {2026},
    note= {Accessed: 2026-04}
}

@misc{dune-analytics,
    title= {Ethereum Uniswap v2 pools TVL},
    author = {Dune},
    howpublished = {\url{https://dune.com/queries/6904458?utm_source=share&utm_medium=copy&utm_campaign=query}},
    year= {2026},
    note= {Accessed: 2026-04}
}

@inproceedings{qin2021empirical,
  title={An empirical study of defi liquidations: Incentives, risks, and instabilities},
  author={Qin, Kaihua and Zhou, Liyi and Gamito, Pablo and Jovanovic, Philipp and Gervais, Arthur},
  booktitle={Proceedings of the 21st ACM internet measurement conference},
  pages={336--350},
  year={2021}
}

@inproceedings{ferreira2024rolling,
  title={Rolling in the shadows: Analyzing the extraction of mev across layer-2 rollups},
  author={Ferreira Torres, Christof and Mamuti, Albin and Weintraub, Ben and Nita-Rotaru, Cristina and Shinde, Shweta},
  booktitle={Proceedings of the 2024 on ACM SIGSAC Conference on Computer and Communications Security},
  pages={2591--2605},
  year={2024}
}

@inproceedings{cernera2023token,
  title={Token spammers, rug pulls, and sniper bots: An analysis of the ecosystem of tokens in ethereum and in the binance smart chain (BNB)},
  author={Cernera, Federico and La Morgia, Massimo and Mei, Alessandro and Sassi, Francesco},
  booktitle={32nd USENIX security symposium (USENIX security 23)},
  pages={3349--3366},
  year={2023}
}

@inproceedings{zhou2021high,
  title={High-frequency trading on decentralized on-chain exchanges},
  author={Zhou, Liyi and Qin, Kaihua and Torres, Christof Ferreira and Le, Duc V and Gervais, Arthur},
  booktitle={2021 IEEE symposium on security and privacy (SP)},
  pages={428--445},
  year={2021},
  organization={IEEE}
}

@misc{tim2025Money,
    author={Craig, Tim},
    title ={Money launderers are mimicking terrible traders to bypass detection, crypto security experts say},
    howpublished={\url{https://www.dlnews.com/articles/defi/lazarus-sandwich-attacks-its-own-trades-to-launder-crypto/}},
    month={March},
    year={2025},
    note={Accessed: 2026-05}
}

@misc{cointelegraph2025crypto,
    author={Cointelegraph},
    title ={Crypto trader gets sandwich attacked in stablecoin swap, loses \$215K},
    howpublished={\url{https://kr.tradingview.com/news/cointelegraph%3A2a602cb81094b%3A0-crypto-trader-gets-sandwich-attacked-in-stablecoin-swap-loses-215k/}},
    month={March},
    year={2025},
    note={Accessed: 2026-05}
}

@misc{cryptoalchemy2025defi,
    author={Crypto\_Alchemy},
    title ={DeFi user loses over \$700K USDC in a sandwich attack},
    howpublished={\url{https://www.binance.com/en/square/post/21489047210377}},
    month={March},
    year={2025},
    note={Accessed: 2026-05}
}

@article{cao2026peb,
  title={PEB Separation and State Migration: Unmasking the New Frontiers of DeFi AML Evasion},
  author={Cao, Yixin and Cheng, Xianfeng and Liu, Yijie},
  journal={arXiv preprint arXiv:2603.26290},
  year={2026}
}

@inproceedings{luo2026light,
  title={Light into Darkness: Demystifying Profit Strategies Throughout the MEV Bot Lifecycle},
  author={Luo, Feng and Li, Zihao and Luo, Wenxuan and He, Zheyuan and Luo, Xiapu and Ma, Zuchao and Song, Shuwei and Chen, Ting},
  booktitle={NDSS},
  year={2026}
}

@misc{PBS-ethereum,
author={Ethereum},
title={Proposer-builder separation},
howpublished={\url{https://ethereum.org/roadmap/pbs/}},
note={Accessed: 2026-05}
}

@inproceedings{park2025frontrunning,
  title={On Frontrunning Risks in Batch-Order Fair Systems for Blockchains},
  author={Park, Eunchan and Yoon, Taeung and Nam, Hocheol and Maram, Deepak and Kang, Min Suk},
  booktitle={Proceedings of the 2025 ACM SIGSAC Conference on Computer and Communications Security},
  pages={918--932},
  year={2025}
}

@inproceedings{mu2024separation,
  author={Mu, Ke and Yin, Bo and Asheralieva, Alia and Wei, Xuetao},
  title={Separation is Good: A Faster Order-Fairness Byzantine Consensus},
  booktitle =	{Proc. NDSS},
  year = {2024},
  pages = {1--17}
}

@inproceedings{kelkar2023themis,
  title={Themis: Fast, strong order-fairness in byzantine consensus},
  author={Kelkar, Mahimna and Deb, Soubhik and Long, Sishan and Juels, Ari and Kannan, Sreeram},
  booktitle={Proceedings of the 2023 acm sigsac conference on computer and communications security},
  pages={475--489},
  year={2023}
}

@inproceedings{mclaughlin2023large,
  title={A large scale study of the ethereum arbitrage ecosystem},
  author={McLaughlin, Robert and Kruegel, Christopher and Vigna, Giovanni},
  booktitle={32nd USENIX Security Symposium (USENIX Security 23)},
  pages={3295--3312},
  year={2023}
}

\appendices
\section{Operational Details of Arbitrage-based \name on Arbitrum}
\label{sec:arbitrum-probing}

As discussed in a previous section~(\S\ref{sec:demo-arbitrum}), executing \name on Arbitrum requires the adversary to place the rate-inflating and arbitrage transactions in separate blocks to mimic an honest arbitrageur, yet execute them as closely as possible to preempt competing arbitrageurs. 
In the remainder of this appendix, we detail how the adversary precisely times its submissions in distinct blocks and reliably outcompetes these rival bots. 

\subsection{Timing Transaction Submissions} 
To mimic an honest arbitrageur's behavior, the adversary should submit the transactions in separate blocks by delaying the arbitrage transaction submissions.
To understand how fast an honest arbitrageur can react to a rate-inflating transaction, we analyze the timing constraints of the Arbitrum network environment.
The Arbitrum sequencer imposes a 200~ms processing delay on every submitted transaction and then publishes the transaction by including it in the nearest block, which is generated every 250~ms.
Consequently, the minimum time for a submitted transaction to be published is greater than 200~ms (e.g., 201~ms), assuming the transaction waits 200~ms in the sequencer and is published as the final transaction in a block.
In other words, if an honest arbitrageur reacts instantly upon observing a rate-inflating transaction in block $i$, their subsequent arbitrage transaction requires some time to process, making it most likely to land in block $i+2$ or later.

Understanding this timing constraint, we formulate a strategy to reliably orchestrate \name transactions by submitting them in separate blocks while keeping the submission delay for the arbitrage transaction as small as possible.
The primary challenge lies in estimating the precise block boundary to accurately time both transaction submissions and minimize the delay between the transactions.
To estimate the block boundaries, we periodically sent probe transactions from an AWS instance in North Virginia and confirmed their final positions within the generated blocks.
Once a probe is confirmed as the final transaction in a block, the adversary calculates the exact submission time for the rate-inflating transaction by adding a target delay ($n \times 250$~ms, where $n$ represents the number of future blocks) to the probe's sending time.
This precise timing allows the rate-inflating transaction to also land as the final transaction of its respective block.
Following this submission, the adversary waits for a delay that is larger than the minimum reaction time for any honest observer (e.g., 201~ms) before submitting the corresponding arbitrage transaction.
We empirically evaluated various delays and present the results in Figure~\ref{fig:arbitrum-success-rate}.
For each delay, we conducted 200 trials and measured the success rate of placing the second transaction exactly two blocks after the first.
Our results demonstrate that a delay of 330~ms achieves a 98.3\% success rate, and this is the delay we used to demonstrate \name in the network.
\begin{figure}[t]
    \centering
    \includegraphics[width=0.5\textwidth]{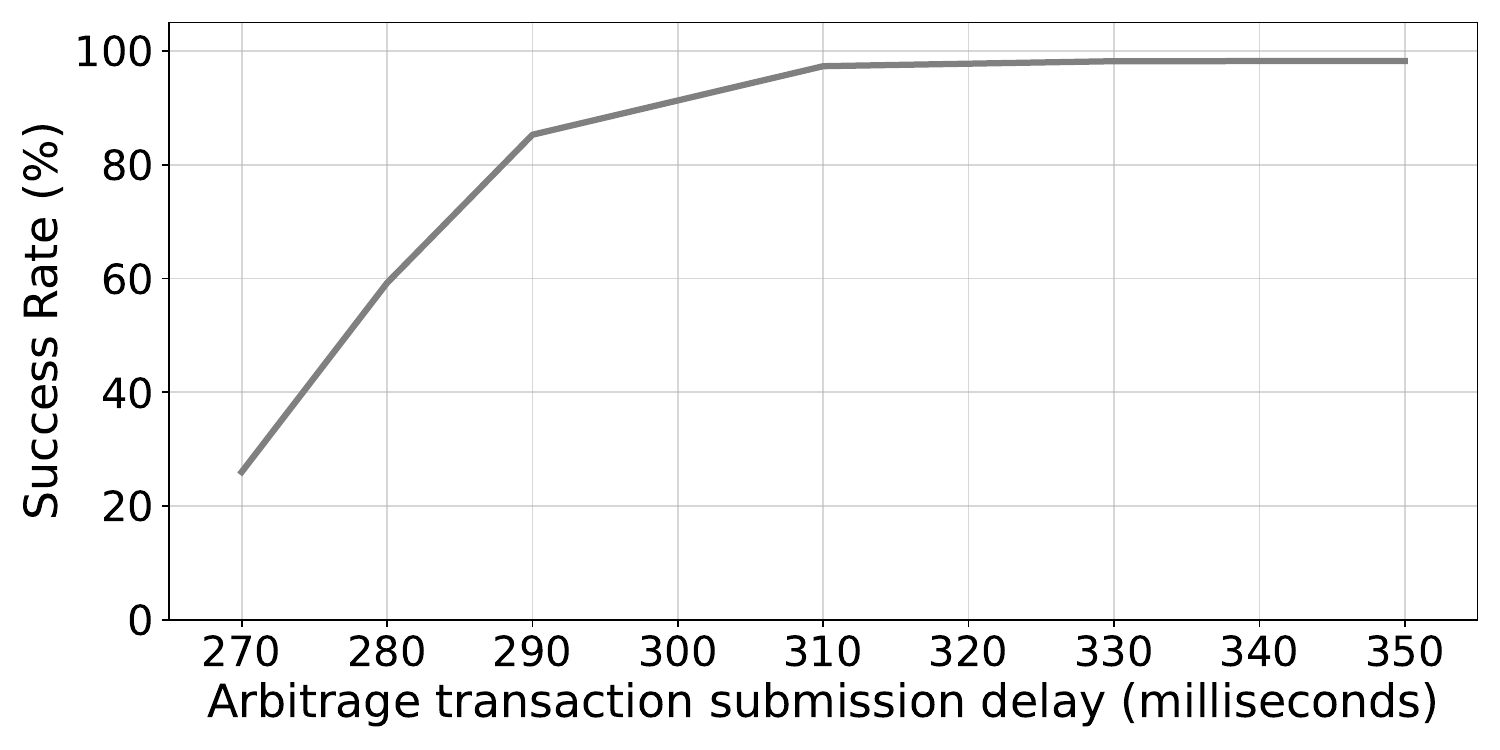}
    \caption{Relationship between success rate and submission gap by threshold}
    \label{fig:arbitrum-success-rate}
\end{figure}

\subsection{Competition Against Other Arbitrageurs}
Having established a reliable method for deniable transaction placement, we now evaluate the adversary's ability to outcompete rival arbitrageurs. 
Because our testnet experiments relied on custom tokens, natural arbitrage competition does not exist; thus, we implemented our own competing arbitrage bot to simulate real-world conditions. 
The competing bot employed an aggressive monitoring strategy: it continuously parsed new blocks for arbitrage opportunities and immediately submitted transactions upon detecting a profitable state.
To maximize their competitiveness, we deployed the bot on AWS instances in Ohio, optimally co-locating them with the testnet sequencer. 
This 100\% success rate stems from an inherent asymmetric advantage: the adversary knows the exact timestamp of the rate-inflating transaction submission, whereas competing bots must opportunistically detect the event only after the block is published. 

\section{Justification on Copula Family Selection}
\label{sec:copula}

As previously discussed, we assign each incident a joint survival probability using a copula to rank the incidents~(\S\ref{sec:multivariate-detection}). 
This appendix briefly discusses the pipeline for calculating this joint survival probability and justifies the selection of copula families.

\subsection{Feature Transformation}

A copula model requires each feature to be expressed as a uniform score in $(0,1)$; raw values cannot be used directly because the four features span incompatible scales and types.
We therefore apply an empirical CDF transform to each feature independently, mapping every observation to its percentile rank within the dataset.
Because the two frequency features are highly discrete (e.g., 60--75\% of incidents share the value 1), they are not suitable to calculate the copula. 
To address this, we randomly add small jitters to tie-break to achieve continuous uniform margins. 
We repeated this process over multiple jitter seeds to confirm that the tie-breaking process has a negligible impact on the final results. 
The resulting uniform scores are then mapped to normal scores via the probit transform for copula fitting.

\subsection{Copula Family Selection}

With the transformed features in hand, we select the copula family that best captures the dependency structure of each dataset.
There are several copula families, each capturing a different dependence structure, and choosing the appropriate one is essential for accurate joint probability estimation. 
We evaluate four candidate families (Gaussian, Clayton, Gumbel, and Frank) using tail dependence plots, Kendall's~$\tau$, and AIC/BIC comparisons. 
Our evaluation reveals that the Clayton, Gumbel, and Frank copulas are unsuitable for our dataset. 
Specifically, the standard 4-dimensional Clayton and Gumbel require positive dependence across all pairs, a condition violated by four of the six feature pairs on both chains; and the standard one-parameter Frank copula cannot capture the heterogeneous pairwise dependence structure observed across the four features.
While the Gaussian copula is the only candidate capable of handling the full dependency structure within a single coherent 4-dimensional model, we must further evaluate its goodness-of-fit for each specific dataset. 

\para{Ethereum.}
The Gaussian copula fits poorly on the Ethereum dataset: only 57\% of the bulk distribution falls within 10\% of its theoretical reference.
This poor fit is caused by strong upper tail dependence between the two frequency features ($\hat{\lambda}_U(0.95) = 0.715$), a characteristic that the Gaussian copula structurally cannot capture.
We therefore fit a \emph{t-copula}, which extends the Gaussian copula with a single degrees-of-freedom parameter~$\nu$ to control tail heaviness.
This approach yields a substantial improvement of $\Delta\mathrm{AIC} = -13{,}397$ over the Gaussian model at the optimal $\nu = 13$. 

\para{Arbitrum.}
The Gaussian copula achieves the best AIC among all valid models for Arbitrum, exhibiting moderate tail dependence across most pairs ($\hat{\lambda}_U(0.95) = 0.216$ for the strongest pair).
This confirms that the Gaussian copula is the appropriate choice for modeling the Arbitrum baseline.

\subsection{Validation}

Having selected the copula family for each chain, we confirm that the fitted models correctly describe their respective datasets. 
To this end, we compare the empirical distance scores against their theoretical reference distributions using Q-Q plots.
Specifically, we use a $\chi$-squared distribution ($\chi^2(4)$) for the Gaussian copula and an F-distribution ($F(4,\nu)$) for the t-copula. 

\para{Ethereum.}
The t-copula demonstrates a strong fit, capturing 85.4\% of the 5th--95th percentile bulk within 10\% of the diagonal, with deviation beginning only at the 99.2nd percentile. 
For comparison, the Gaussian copula captures only 57\% of the bulk and begins deviating at the 94.6th percentile.
This contrast confirms that the t-copula is better suited for the Ethereum dataset.

\para{Arbitrum.}
The Gaussian copula captures 85.3\% of the bulk within 10\% of the diagonal, with deviation beginning at the 96.2nd percentile. 
This outcome confirms that the Gaussian model provides a good fit for Arbitrum.
\end{document}